\documentclass[aps,prl,twocolumn,showpacs, notitlepage]{revtex4-1}
\usepackage{graphicx}% Include figure files
\usepackage{hyphenat}
\begin{document}

\title{Hydrostatic and uniaxial pressure tuning of iron-based superconductors: Insights into superconductivity, magnetism, nematicity and collapsed tetragonal transitions}

\author{Elena Gati$^{1,2}$}
\author{Li Xiang$^{1,2}$}
\author{Sergey L. Bud'ko$^{1,2}$}
\author{Paul C. Canfield$^{1,2}$}
\address{$^{1}$Ames Laboratory, US Department of Energy, Iowa State University, Ames, Iowa 50011, USA}
\address{$^{2}$Department of Physics and Astronomy, Iowa State University, Ames, Iowa 50011, USA}

\begin{abstract}
Iron-based superconductors are well-known for their intriguing phase diagrams, which manifest a complex interplay of electronic, magnetic and structural degrees of freedom. Among the phase transitions observed are superconducting, magnetic, and several types of structural transitions, including a tetragonal-to-orthorhombic and a collapsed-tetragonal transition. In particular, the widely-observed tetragonal-to-orthorhombic transition is believed to be a result of an electronic order that is coupled to the crystalline lattice and is, thus, referred to as nematic transition. Nematicity is therefore a prominent feature of these materials, which signals the importance of the coupling of electronic and lattice properties. Correspondingly, these systems are particularly susceptible to tuning via pressure (hydrostatic, uniaxial, or some combination). We review efforts to probe the phase diagrams of pressure-tuned iron-based superconductors, with a strong focus on our own recent insights into the phase diagrams of several members of this material class under hydrostatic pressure. These studies on FeSe, Ba(Fe$_{1-x}$Co$_x$)$_2$As$_2$, Ca(Fe$_{1-x}$Co$_x$)$_2$As$_2$ and CaK(Fe$_{1-x}$Ni$_x$)$_4$As$_4$ were, to a significant extent, made possible by advances of what measurements can be adapted to the use under differing pressure environments. We point out the potential impact of these tools for the study of the wider class of strongly correlated electron systems. 
\end{abstract}

\maketitle

\section{Introduction}

During the dozen years since the discovery of superconductivity in the iron-based material LaFeAsO$_{1-x}$F$_x$ \cite{Kamihara06}, the class of iron-based superconductors has become an important platform for the development of a microscopic understanding of unconventional superconductivity in strongly correlated electron systems \cite{Paglione10,Johnston10,Canfield10,Mazin10,Hosono15,Hosono18}. In general, the competing tendencies \cite{Imada98} towards different ground states in strongly correlated electron systems are believed to be at the origin of their complex phase diagrams \cite{Canfield16}, in which a variety of intriguing phases, such as superconductivity, magnetism, orbital and structural orders, are often found in close proximity \cite{Lee06,Steglich13,Toyota07,Uemura09}. The tuning of correlated electron systems is essential to explore their rich phase diagrams and to induce phase transitions into novel states. The most common ways to tune materials in laboratory experiments involve either chemical substitutions, leading to changes of the crystallographic lattice parameters (referred to as chemical pressure) and often also the band filling via doping, or the application of physical pressure. Physical pressure as a tuning parameter modifies crystallographic lattice parameters, that in turn induce changes in the electronic properties. As such, in contrast to chemical substitution, tuning by physical pressure does not involve changing levels of disorder. Disorder is known to complicate the analyses of electronic phenomena \cite{Nie14}, since any level of disorder acts as a perturbation, which might tip the balance between the various, almost degenerate electronic states. Iron-based superconductors can be considered particularly suited to pressure studies, since the presence of various types of structural orders in proximity to electronic and magnetic orders in the phase diagrams indicates a prominent interplay of electronic, magnetic, and lattice degrees of freedom \cite{Fernandes14,Canfield09b}. Here, we review recent efforts to tune and probe different phases in iron-based superconductors by hydrostatic as well as uniaxial pressure, with a strong focus on our own work on various members under hydrostatic pressure. In doing so, we also outline how recent advances in experimental techniques have lead to an improved understanding of phases and their interplay in pressure-tuned iron-based superconductors. 

		\begin{figure*}[t]%
		\begin{center}
		\includegraphics*[width=0.9\textwidth]{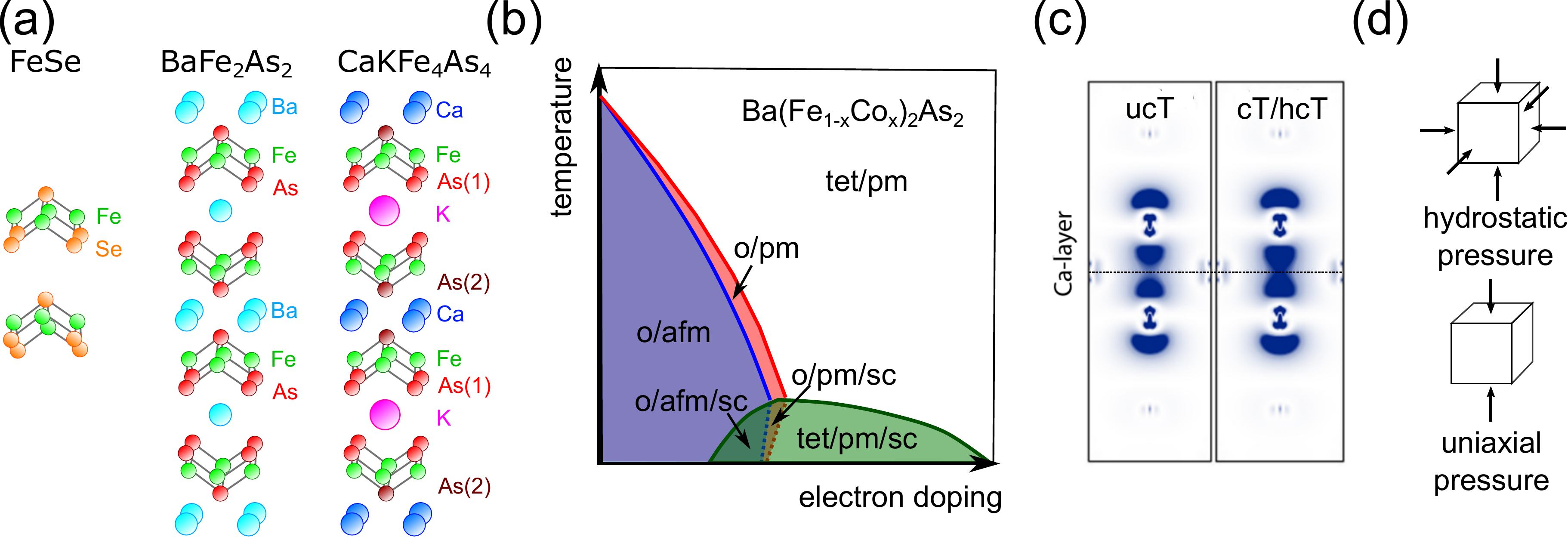}
		\caption{Crystal structures, the archetypal phase diagram, the collapsed-tetragonal transition and tuning parameters of iron-based superconductors. (a) Crystallographic structure of representative members of the 11 family (FeSe), the 122 family (BaFe$_2$As$_2$) and the 1144 (CaKFe$_4$As$_4$) family \cite{Hsu08,Rotter08,Iyo16}; (b) Phase diagram of BaFe$_2$As$_2$ upon electron doping, which is considered as a typical phase diagram for iron-based superconductors; tet stands for tetragonal, pm for paramagnetic, o for orthorhombic, afm for antiferromagnetic and sc for superconducting \cite{Canfield10}; (c) Non-spin-polarized electron density in the $ac$ plane associated with the As $p_z$ orbitals above and below the Ca-plane in the uncollapsed tetragonal (ucT) and either collapsed tetragonal (cT) or half collapsed tetragaonal (hcT) state in either CaFe$_2$As$_2$ or CaKFe$_4$As$_4$, respectively (after \cite{Kaluarachchi17,Borisov18,Borisov19}); (d) Schematic view on the forces on the surfaces of a sample, which are associated with hydrostatic (top) and uniaxial (bottom) pressure on a crystal lattice (which, for simplicity, is represented by the cube).}
		\label{fig:overviewfigure}
		\end{center}
		\end{figure*}

\textit{Crystallographic structure -} The common structural motif in the layered iron-based superconductors is the layer of iron atoms, which form a square lattice and are tetrahedrally coordinated by pnictogen or chalcogen atoms (see Fig.\ref{fig:overviewfigure}\,(a)) \cite{Paglione10,Iyo16}. The resulting trilayers are the building blocks for Fe-based superconductors and can be stacked with or without spacing layers, depending on the specific compound. Different iron-based superconductors are classified according to their stochiometry. For example, in the ''11'' materials, such as FeSe, the trilayers are stacked along the $c$ axis without any intermediate spacing layers (see Fig.\ref{fig:overviewfigure}\,(a)). In contrast, in the ''122'' $A$Fe$_2$As$_2$ structure, which belongs to the long-known ThCr$_2$Si$_2$ structure, the trilayers alternate along the $c$ axis with layers consisting of alkali or alkaline-earth metals, as shown in Fig.\ref{fig:overviewfigure}\,(a). Other systems with a ''111'' and ''1111'' structure type (e.g. LiFeAs \cite{Tapp08} and LaFeAsO \cite{Kamihara06}) are also well known and have been investigated since the early days. In 2016, Iyo \textit{et al.} \cite{Iyo16} discovered that a new class of iron-based superconductors with a ''1144'' stochiometry is formed in Ca$A$Fe$_4$As$_4$ with $A\,=\,$K, Rb, Cs (see Fig.\ref{fig:overviewfigure}\,(a)). Owing to the sizable differences in the ionic radii of the Ca and $A$ atoms, respectively, this structure is characterized by alternating Ca and $A$ layers, separated by the FeAs trilayers. In contrast to the solid solution (Ba$_{1-x}A_x$)Fe$_2$As$_2$ \cite{Rotter08}, where the Ba and $A$ ions occupy randomly a single site, the 1144 compounds are well-ordered line compounds as a result of the layer-by-layer segregation of the Ca and $A$ ions.

\textit{Phase diagrams: Superconductivity, magnetism and nematicity -} Given the large chemical variety and, at the same time, the availability of high-quality, large-sized single crystals of members of the 122 family, investigations on these materials have significantly shaped the canonical picture of the phase diagram of iron-based superconductors (see Fig.\ref{fig:overviewfigure}\,(b) for the phase diagram of electron-doped BaFe$_2$As$_2$ \cite{Canfield10,Avci12}). The parent 122 compounds are tetragonal and paramagnetic at high temperatures and, like many other unconventional superconductors (see Ref. \cite{Paglione10} and Refs. cited therein), undergo a transition to an antiferromagnetic state upon cooling at $T_{N}$. This antiferromagnetic order is in most cases a stripe-type order \cite{Huang08}, as a consequence of which the two in-plane directions become unequal. Correspondingly, the magnetic transition is intimately coupled to a structural phase transition at $T_s$ \cite{Canfield10}, at which the tetragonal, $C_4$, crystal symmetry is reduced to an orthorhombic, $C_2$ symmetry. In some systems, like CaFe$_2$As$_2$ \cite{Ni08b} or (Ba$_{1-x}$K$_x$)Fe$_2$As$_2$ \cite{Avci12}, the magnetic and structural transition occur simultaneously ($T_N\,=\,T_s$), and are first-order transitions, whereas in other systems, like Ba(Fe$_{1-x}$Co$_x$)$_2$As$_2$, the structural transition is found to precede the magnetic transition ($T_N\,<\,T_s$), with both transitions being second-order transitions \cite{Canfield10}. Upon suppression of this magneto-structural order to sufficiently low temperatures by a suitable tuning parameter, like doping \cite{Canfield10}, isovalent substitution \cite{Thaler10,Boehmer12} on different crystallographic sites or physical pressure \cite{Colombier09,Alireza08}, superconductivity emerges. The superconducting critical temperature $T_c$ often depends on the tuning parameter in such a way that a dome of superconductivity forms in the phase diagram, with maximum $T_c$ located in close proximity to where the magnetic and structural phase lines from the normal state extrapolate to zero Kelvin. Superconductivity and magnetic-orthorhombic order are believed to compete with each other, as, e.g., indicated by a decrease of magnetic-structural order parameters as well as a break and even back-bending of the phase line(s) when superconductivity sets in \cite{Nandi10}. 

The proximity of superconductivity and magnetism has sparked ideas of a magnetically-driven mechanism of superconductivity \cite{Mazin10}. However, to unravel the superconducting mechanism \cite{Hirschfeld11,Chubukov12,Si16}, it is important to consider all salient ground states and their respective electronic fluctuations. In this regard, an understanding of the origin of the structural phase transition has become a central theme in the research on iron-based superconductors \cite{Fernandes14}. By now, it is well established that this transition is not an ordinary structural transition, which is driven by lattice degrees of freedom. Instead, it is widely believed that the structural transition is driven by electronic degrees of freedom \cite{Fernandes14,Chu10,Chu12,Tanatar10,Nandi10,Boehmer14}, and as such is intimately related to the same degrees of freedom that are responsible for superconductivity and magnetism \cite{Fernandes12b}. Based on an analogy to liquid crystals, the orthorhombic state, which is characterized by an in-plane anisotropy, associated with a reduced symmetry compared to the high-temperature tetragonal state, is commonly referred to as a ''nematic'' state \cite{Fernandes14,Fernandes19,Fradkin10,Fradkin15}. A key question in literature relates to the primary order parameter and thus, the microscopic origin of nematicity \cite{Fang08,Xu08,Si08,Fernandes12c,Capati11,Wysocki11,Hu12,Krueger09,Lee09,Yin10,Lv10,Chen10,Applegate12}. Both, spin as well as orbital degrees of freedom are considered as promising candidates for the driving force behind nematicity. Given that both of these types of order are known to create an in-plane anisotropy and are coupled to each other, the identification of the driving mechanism turns out be particularly complicated \cite{Fernandes14} (''chicken-or-egg'' problem). This dilemma has led to intensive research efforts on other iron-based superconductors, which display more unusual apparent relations of magnetism, nematicity and superconductivity. For example, two extreme cases of an remarkable interplay of nematicity, magnetism and superconductivity are given by FeSe \cite{Boehmer17rev} and CaKFe$_4$As$_4$ \cite{Iyo16,Meier16}. Both of this materials are superconductors in their parent form with $T_c\,\approx\,$8\,K and 35\,K, respectively. Concerning their magnetic and structural properties, FeSe has received and continues to receive a lot of attention, since it displays nematic order at moderate temperatures below $T_s\,\approx\,90\,$K at ambient pressure, but lacks any magnetic order down to lowest temperatures. It therefore represents a unique and promising example case to study the physics of a purely nematic state, and its interrelation with superconductivity \cite{Boehmer17rev} (see Sec.\,\ref{sec:FeSe}). In contrast, as we will discuss below in Sec.\,\ref{sec:CaK1144} in more detail, CaKFe$_4$As$_4$ is located in the proximity of a new type of magnetically-ordered state, the so-called hedgehog spin-vortex order, which does not break the tetragonal symmetry of the high-temperature state \cite{Meier18}. This magnetic order was found to be stabilized by Ni-substitution on the Fe site (hole doping). Correspondingly, it was suggested that the series of CaK(Fe$_{1-x}$Ni$_x$)$_4$As$_4$ allows for the study of the impact of magnetic fluctuations on superconductivity in the absence of nematicity \cite{Ding18}.

Overall, the prominent interplay of electronic, magnetic and structural degrees of freedom makes the iron-based superconductors particularly amenable to the tuning via physical pressure. As we will discuss in detail below, moderate pressures are sufficient to tune these materials through different ground states. For example, magnetic order can be induced in FeSe by the application of moderate hydrostatic pressures of $p\,\approx\,0.9$\,GPa \cite{Bendele10}. Specifically, we will describe how pressure tuning of selected systems has enabled us to gain new insights about the nature and the mutual interplay of superconductivity, magnetism and nematicity, driven by the recent advancements of experimental techniques.

\textit{Collapsed-tetragonal transitions -} In addition to the previously mentioned electronic and structural phase transitions, another structural instability is well known for those systems crystallizing in the ThCr$_2$Si$_2$ structure \cite{Hoffmann85,Huhnt98,Bishop10,Yu15b,Naumov17,Drachuck17}, such as the $A$Fe$_2$As$_2$ and $Ae$Fe$_2$As$_2$ ($A\,=\,$alkali and $Ae\,=\,$alkaline-earth metal) systems \cite{Canfield09b,Kasinathan11,Uhoya10}. The formation of $p_z$ bonds in case of a sufficiently short As interlayer-distance \cite{Hoffmann85} (see Fig.\ref{fig:overviewfigure}\,(c) for the non-polarized electron density across the collapsed Ca layer in CaKFe$_4$As$_4$) results in a structural phase transition from the regular, uncollapsed tetragonal structure (tet) to the so-called collapsed tetragonal (cT) structure. This structural transition is associated with a drastic shrinkage of the $c$ axis lattice parameter and an expansion of the $a$ axis lattice parameter. In CaFe$_2$As$_2$ and related systems, these lattice parameter changes are accompanied by significant changes of the electronic properties \cite{Kreyssig08,Canfield09,Ran12,Gati12,Furukawa14}, related to the underlying changes of the dimensionality of the electronic band structure and changes of the Fe magnetic moment \cite{Yildrim09,Borisov18}, which can lead, e.g., to a loss of superconductivity or magnetism. Thus, the manipulation of the pnictogen interlayer-distance can be used as a tool to investigate the response of the electronic properties to this change of crystal structure. The use of physical stress allows a direct manipulation of this distance, and thus is a very suitable tuning parameter for the investigation of this structural collapse and the associated electronic changes. We will discuss how pressure tuning different iron-pnictides through critical interlayer distances for pnictogen bondings has been important for (i) inferring crucial ingredients for the appearance of superconductivity, (ii) the discovery of a new type of a collapsed transition, the so-called half-collapsed tetragonal transition and (iii) the discovery of a superelastic behavior in intermetallic compounds with exceptionally large and recoverable strain.

\textit{Hydrostatic vs. uniaxial pressure -} Whereas we have introduced above why iron-based superconductors are promising candidate systems for exploring the effects of pressure, it is our aim to also outline here how different types of pressures, which are experimentally available, affect the electronic properties of this material class. In more detail, we will emphasize why specific ground states in iron-based superconductors, such as nematicity, have motivated the use of different types of physical pressure and thus, how the research on iron-based superconductors contributed significantly to the development of new or advanced experimental methods, which are of relevance for the wider class of correlated electron materials. Specifically, we will focus on the impact of hydrostatic and uniaxial pressure (see Fig.\ref{fig:overviewfigure}\,(d)), which are distinct in their effect on the underlying crystal lattice \cite{Hicks14b}. Whereas for hydrostatic pressure the force is equally distributed to all crystal surfaces, and thus the tuning parameter of hydrostatic pressure itself is non-directional, uniaxial pressure is highly directional, since the force is applied along a specific crystallographic direction. Correspondingly, the comparison of hydrostatic vs. uniaxial pressure allows for the investigation of how interesting electronic orders responds to distinctly different lattice deformations.

\textit{Overview -} The remainder of this article is structured as follows. In section \ref{sec:experimentalmethods}, we will describe experimental methods to apply hydrostatic and uniaxial pressure, and outline advances in experimental techniques, which allow for a determination of the phase diagrams under pressure. In Secs.\,\ref{sec:thermodynamicphasediagrams} and \ref{sec:Casystems-cT}, we will then describe our own efforts in the determination and refinement of the temperature-pressure phase diagrams of several members of the family of iron-pnictides using hydrostatic pressure. In particular, we will focus on recent insights into the interplay of superconductivity, nematicity and magnetism under hydrostatic pressure in FeSe (Sec.\,\ref{sec:FeSe}) and Ba(Fe$_{1-x}$Co$_x$)$_2$As$_2$ (Sec.\,\ref{sec:Ba122}). We will also discuss the occurrence of structural instabilities, i.e., collapsed-tetragonal transitions, related to the interlayer bonding of the pnictogen atoms, under hydrostatic pressure and their effect on the magnetic and superconducting properties, in Ca(Fe$_{1-x}$Co$_x$)$_2$As$_2$ (Sec.\,\ref{sec:Ca122}) and CaK(Fe$_{1-x}$Ni$_x$)$_4$As$_4$ (Sec.\,\ref{sec:CaK1144}). After these discussions on the effect of hydrostatic pressure, we will shortly outline the role of uniaxial pressure for the tuning of the collapsed-tetragonal transition in Sec.\,\ref{sec:superelasticity}. Following this, we will summarize in Sec.\,\ref{sec:uniaxialstrain} the current understanding of the role of uniaxial pressure for probing and tuning the magnetic, structural and superconducting properties of iron-based superconductors. Afterwards, we will show in Sec.\,\ref{sec:hydrostaticanduniaxial} our recent efforts, which allow for the study of the combined effects of hydrostatic and uniaxial strain, and outline their potential for the study of the iron-based superconductors. We conclude this paper in Sec.\,\ref{sec:summary} by providing a summary and outlook, which highlights, how the improved set of techniques, now available under pressure, might be relevant for the study of the wider class of correlated materials.

\section{Tuning by hydrostatic and uniaxial pressure: Experimental methods}
\label{sec:experimentalmethods}

In this section, we will describe experimental methods to apply physical pressure to correlated electron systems. In particular, we will review methods of how to apply hydrostatic as well as uniaxial pressure. This will be followed by a summary of recent advances in experimental measurements, that have been adapted to these pressure environments so as to detect the properties of correlated matter under pressure. 

Prior to a detailed description of these experimental methods, we want to introduce the notion of ''stress control'' vs. ''strain control'' (see Hicks \textit{et al.} \cite{Hicks14} and Barber \textit{et al.}  \cite{Barber19}). This clarification is needed, since either stress or strain corresponds to the control variable depending on the experimental design, despite the fact that an applied stress induces a strain in a material and vice versa. In short, whether stress or strain is controlled in a specific setup depends on the spring constant of the apparatus $k_{app}$ compared to the sample's spring constant $k_s$. For a small spring constant of the apparatus $k_{app}\,\ll\,k_s$, the control parameter is stress, whereas for a large spring constant of the apparatus $k_{app}\,\gg\,k_s$, strain is the control parameter. As suggested by the above analysis \cite{Hicks14,Barber19}, we will label the different techniques by their control parameter. In addition, for the special case of uniaxial strain as a control parameter, it is important to note that the application of uniaxial strain along a particular axis will also result in straining the sample along the other crystallographic directions. This effect is known as Poisson's effect, and the so-called Poisson ratio $\nu$ measures the ratio of the strains induced along different crystallographic directions. Typically, for ordinary materials, $|\nu|\, < 1$ \cite{Gere97}.

\textit{Hydrostatic stress -} Moderate hydrostatic pressures are typically applied by placing a sample into the small sample space of a pressure cell \cite{Eremets96,Fujiwara80,Budko84,Bridgman52}, in which the sample is then surrounded by a pressure-transmitting medium. The medium ensures that the applied force, which results from, e.g., the application of a force to a piston or an anvil, is equally distributed to all sample surfaces. Note that in these kinds of experiments, the spring constant of the pressure medium (rather than the one of the pressure cell body etc.) is the one, which is typically a lot smaller than that of the sample ($k_{app}\,<\,k_s$), and thus, stress and not strain is the control parameter (Note that this inequality causes concerns and puts some restrictions on the choice of the pressure medium, e.g., for organic samples).

The degree of hydrostaticity of the applied pressure depends very sensitively on the properties of the chosen medium, such as the solidification temperature. The solidification of the pressure medium typically leads to deviations from the desired ideal hydrostaticity \cite{Torikachvili15}. For this reason, the use of Helium gas, which is either in its gaseous or liquid state down to much lower temperatures \cite{Keesom26,Pinceaux79} compared to any liquid pressure medium \cite{Torikachvili15} (liquid pressure medium denotes that the medium is liquid at room temperature and ambient pressure), ensures the best hydrostatic conditions at low temperatures. However, the use of He gas is frequently limited to the low-pressure range $(p \lesssim 1\,$GPa). Larger pressures ($p \lesssim 10\,$GPa) are often achieved by using a liquid pressure medium, which is chosen such that it does not solidify at room temperature across the full pressure range of interest. In this way, hydrostaticity of the medium is ensured during the pressure change, which is performed at room temperature using a hydraulic press. Conversely, given that Helium gas only solidifies at very low temperatures, the use of Helium gas as a pressure transmitting medium also provides the appealing opportunity to perform isothermal measurements as a function of pressure, i.e., pressure can be varied \textit{in situ} at constant (low) temperature \cite{Gati16}. 

\begin{figure}[t]%
		\begin{center}
		\includegraphics*[width=\linewidth]{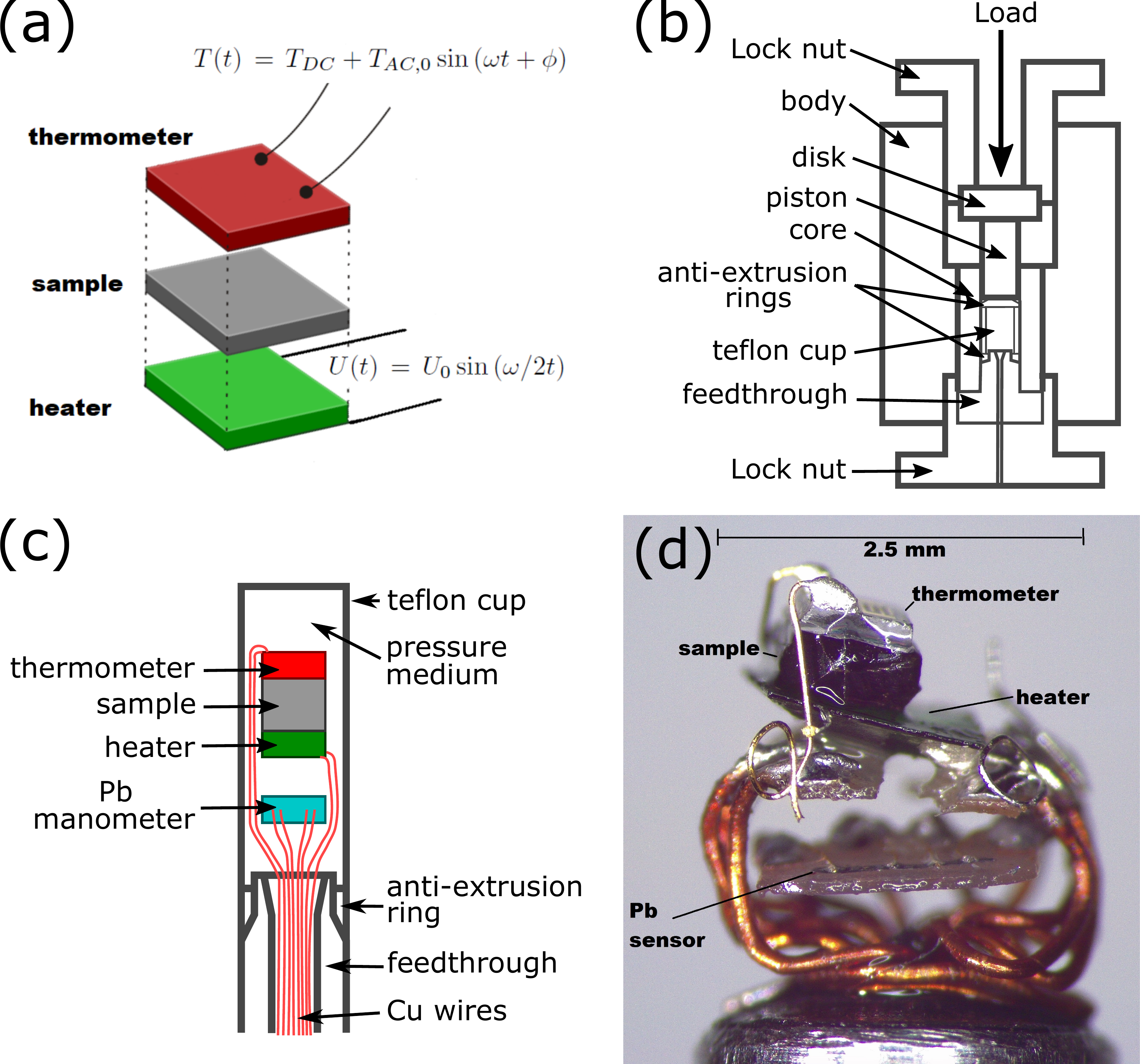}
		\caption{Specific heat under hydrostatic pressure in piston-pressure cells. (a) Schematic view of the arrangement of the heater, sample and thermometer for $ac$ specific heat measurements under pressure; (b) Schematic diagram of the piston-pressure cell which is used in most of our own work, which are presented in this manuscript; (c) Enlarged view on the sample assembly inside the sample space (inside the Teflon cup) and the pressure-cell feedthrough; (d) Photograph of the sample assembly, and the Pb manometer, on top of the pressure-cell feedthrough. Reprinted with permission from Ref. \cite{Gati19a}, Copyright AIP Publishing 2019.}
		\label{fig:experimentalsetup}
		\end{center}
		\end{figure}

In this paper, we will review results from measurements, performed in Helium-gas pressure cells $(p \lesssim 0.3\,$GPa) \cite{Unipress}, piston-pressure cells $(p \lesssim 2.5\,$GPa) (see Fig.\,\ref{fig:experimentalsetup}\,(b)) \cite{Budko84} with a 4:6 mixture of light mineral oil and $n$-pentane as a pressure-transmitting medium as well as modified Bridgman anvil cells $(p \lesssim 6\,$GPa) \cite{Colombier07} with a 1:1 mixture of $n$-pentane:iso-pentane as a pressure-transmitting medium. We will discuss the implications of potential non-hydrostatic pressure components, whenever appropiate. Also, we will refer to liquid-medium pressure cells, whenever the medium is liquid at room temperature, whereas we refer to gas-medium cells, when the medium is gaseous at room temperature. The absolute pressure value at low temperatures is typically determined by the shift of the superconducting transition temperature of either lead (Pb) \cite{Eiling81}, tin (Sn) \cite{Smith67} or indium (In) \cite{Jennings58}.

\textit{Uniaxial stress and strain -} Uniaxial pressure is distinct from hydrostatic pressure, as the applied force, which acts on the crystal, is highly directional. Compressive uniaxial stress or strain can, e.g., be experimentally realized by fixing a sample between two anvils \cite{Pfleiderer97,Osterman85}. Conversely, tensile stress or strain can be achieved by pulling on two ends of a sample \cite{Cooks97,Brandt80}, e.g. in so-called ''quartz puller'' \cite{Cooks97} or ''horseshoe'' devices \cite{Tanatar10}. More recent technical developments involve the use of piezoelectric actuators \cite{Shayegan03,Chu12,Kuo13,Kuo16,Hicks14,Barber19}, which can be conveniently, controllably and continuously strained by the application of an external voltage and therefore allow for a control of uniaxial strain or stress of samples \textit{in situ} at low temperatures. To this end, samples are either directly attached to the actuator \cite{Chu12,Kuo13,Kuo16}, or placed between two plates \cite{Hicks14,Barber19}, one of which can be moved by the piezoelectric actuator. In each case, care has to be taken to ensure a homogeneous strain and stress distribution across the samples and a non-ideal strain and stress transmission of the glue (epoxys), which is used to fix the samples ridigly to the apparatus, must be taken into account \cite{Hicks14,Kuo16}. In addition, the thermal expansion mismatch between the sample and the device inevitably leads to temperature-dependent changes of the absolute strain, the samples are exposed to \cite{Hicks14,Kuo16}. In novel piezo-based uniaxial-stress cell designs \cite{Hicks14,Barber19}, which are commercially available from Razorbill Instruments \cite{Razorbill}, the expansion mismatch effects were minimized by placing three actuators in series, which effectively cancels the apparatus' temperature-induced thermal expansion effects on the sample. The amount of applied strain can be inferred, to varying degrees of success (i.e. depending on relative $k_{app}$ and $k_s$ values) via measurements of the resistances of strain gauges, or via capacitance measurements of a plate capacitor. In even more recent designs of piezo-based devices \cite{Barber19}, the apparatus has been extended to house a force sensor in addition to the displacement sensor. The combination of both is advantageous for the detection of potential non-elastic deformations of either the sample or the sample mount. 

\textit{Measurement probes under pressure -} For the determination of phase diagrams of correlated materials at ambient pressure, a combination of thermodynamic and transport measurements is usually employed. Among the large number of techniques, resistance, magnetization and specific heat measurements are frequently used tools to detect phase transitions and determine their transition temperatures. Given the limited space in pressure cells and the large amount of pressure cell material, different aspects, specific to the measurement probe, need to be considered when performing resistance, magnetization or specific heat measurements under hydrostatic pressure. Lab-built (see e.g.\,\cite{Budko84,Colombier07}), as well as commercially-available pressure cells up to pressures as high as several GPa are available, which guarantee an electrical connection into the sample space, see e.g. Almax Ltd. \cite{Almax}. In many cases, these cells are then used for resistance measurements by employing a standard four-point configuration. For magnetization measurements under pressure, commercial pressure cells (such as the HMD High Pressure Cell for Magnetometry, sold by Quantum Design \cite{QuantumDesign}) are available, which allow measurements up to 1 GPa in commercial Quantum Design MPMS magnetometers. A big challenge here is to subtract the sizable background contribution of the pressure cell from the measured signal \cite{Coak20}. Alternatively, measurements of the magnetic susceptibility by using an $ac$ technique can be performed inside a pressure cell. Certainly, this technique is not as sensitive to small moments as the commercial SQUID-based MPMS, but it offers the possibility to measure frequency-dependent magnetic properties, which can be of relevance for, e.g., spin glasses \cite{Loehneysen78} and has the great potential for significant miniaturization, which is essential for application at even higher pressures \cite{Braithwaite07,Hamlin07}. Specific heat measurements under pressure turn out to be particularly difficult, as (i) the sample mass is usually very limited and (ii) the heat flow through sample, medium etc. is difficult to account for in modelling of the temperature relaxation, which leads to complications in the subtraction of large background contributions from the measured data. In this regard, the technique of $ac$ calorimetry has proven to be particularly suited \cite{Eichler79,Bonilla74,Baloga77,Chen93,Bouquet00,Wilhelm03,Kubota08,Umeo17}. Here, the sample is heated by an oscillatory heat source and the resulting temperature oscillation contains information on the specific heat of the sample (see Fig.\,\ref{fig:experimentalsetup}\,(a)). The main advantage for the use of this technique in the pressure cell is related to the choice of the heating frequency. This allows for the performance of measurements on a time scale, which is much faster than the relaxation time to the bath (i.e., the pressure medium and the pressure cell). As a result, to a first approximation, the sample is effectively decoupled from the bath, which in principle allows for the extraction of absolute values of the specific heat on a semi-quantitative level \cite{Eichler79}. Although $ac$ calorimetry measurements under pressure have a long-standing history in the community \cite{Eichler79,Bonilla74,Baloga77,Chen93,Bouquet00,Wilhelm03,Kubota08,Umeo17}, its use was typically restricted to narrow temperature ranges due to reasons related to the sensitivity of the thermometers used. In our efforts to determine the phase diagram of iron-based superconductors under hydrostatic pressure, which typically undergo a cascade of phase transitions over wide temperature ranges, we recently reported on an optimization of the thermometry of such an $ac$ calorimetry setup to measure specific heat over wide temperature ranges in conventional piston-pressure cells up to 2.5 GPa (see Fig.\,\ref{fig:experimentalsetup}) \cite{Gati19a}. By utilizing commercially-available Cernox thermometers to pick up the temperature oscillations of the sample, we demonstrated that we can measure the specific heat of a sample of interest over a temperature range as wide as up to 150\,K  (and likely even larger). In Sec. \ref{sec:thermodynamicphasediagrams} and \ref{sec:Casystems-cT} below, we will show how our combined efforts of transport and thermodynamic investigations under hydrostatic pressure (in particular those of the specific heat, using the aforementioned optimized $ac$ calorimetric technique) have advanced the understanding of the phase diagrams of several iron-based superconductors. For completeness, we would also like to mention that efforts have been made to adapt other thermodynamic probes into pressure cells. Specifically, we want to refer the reader to the successful implementation of a capacitive dilatometer into a Helium gas pressure cell, which allows for high-resolution thermal expansion measurements of solids up to 0.25\,GPa \cite{Manna12,Gati16}.

For thermodynamic and transport measurements under uniaxial stress/strain, similar challenges have to be faced as is the case for measurements under hydrostatic pressure. In particular, to measure specific heat, the $ac$ calorimetry technique has been succesfully employed to e.g. measure the specific heat signature of the superconducting transition in Sr$_2$RuO$_4$ under uniaxial strain \cite{Li19}. The choice in favor of the technique of $ac$ calorimetry is also motivated by the desire to effectively decouple the sample from the uniaxial strain cell. Conversely, given the recent successes in tuning uniaxial strain \textit{in situ}, this has initialized ideas to use $ac$ elastocaloric measurements as a tool to explore specific heat \cite{Ikeda19}. The key idea here is, that similar to the oscillating heat in $ac$ calorimetric experiments, an oscillating strain can induce a temperature oscillation, related to the specific heat of the sample and which can be recorded by a thermometer. Proof-of-principle tests of this idea were presented for the iron-pnictide BaFe$_2$As$_2$ \cite{Ikeda19}. The extension to use the $ac$ elastocaloric technique for measurements of the specific heat at finite offset strains is underway \cite{HicksFisher}. 

\section{Effect of hydrostatic pressure on superconductivity, magnetism and nematicity}
\label{sec:thermodynamicphasediagrams}

In the following, we will describe the current understanding of the temperature-pressure phase diagrams of selected iron-based superconductors. In this section, we will primarily focus on the prominent interplay of superconductivity, nematicity, magnetism in FeSe and Ba(Fe$_{1-x}$Co$_x$)$_2$As$_2$ under hydrostatic pressure. These systems are, for different reasons, often referred to in discussions of the universal picture of the phase diagrams of iron-based superconductors. Although our studies of the collapsed tetragonal (cT) phase could logically fit into this section, the cT is different enough that we discuss it separately in section \ref{sec:Casystems-cT} of this paper.

	\subsection{Phase diagram of FeSe under hydrostatic pressure}
	\label{sec:FeSe}
	
	In FeSe, the absence of long-range magnetic order despite the presence of nematic order \cite{McQueen09} is striking and resulted in a large interest \cite{Boehmer17rev} in this purely nematic state and its interrelation with superconductivity. Given that strong magnetic fluctuations were observed at ambient pressure \cite{Boehmer15,Wang16,Wang16b,Kotegawa12,Rahn15,Ma17}, the ''sought-for'' magnetically-ordered ground state was found to be stabilized in FeSe by the application of modest hydrostatic pressure, $p\,\gtrsim\,0.9\,$GPa, as first demonstrated by $\mu$SR measurements under pressure \cite{Bendele10}. This observation initiated ideas that the phase diagram of FeSe might depict features which are compatible with the universal phase diagram of iron-based superconductors \cite{Wang16c,Kothapalli16} and therefore might still be consistent with a magnetically-driven mechanism for nematicity and superconductivity. However, this picture continues to be questioned, in particular for the low-pressure non-magnetic nematic state \cite{Khasanov18,Boehmer19}, for which an orbital-driven mechanism is discussed on equal footings to the spin-driven one, see e.g.\cite{Chubukov16,Glasbrenner15,Onari16,Yamakawa16,Chubukov15,Yu15,Wang15,Scherer17}. At the same time, hydrostatic pressure as a tuning knob in FeSe is particularly interesting, since the application of hydrostatic pressure also revealed a large tunability of the superconducting $T_c$, i.e., an increase of $T_c$ by almost a factor of 4 by a pressure increase of $\approx\,4$\,GPa \cite{Mizuguchi08,Medvedev09,Margadonna09,Garbarino09,Miyoshi09,Okabe10,Kaluarachchi16,Terashima16}.
	
		\begin{figure}[t]%
		\begin{center}
		\includegraphics*[width=0.9\linewidth]{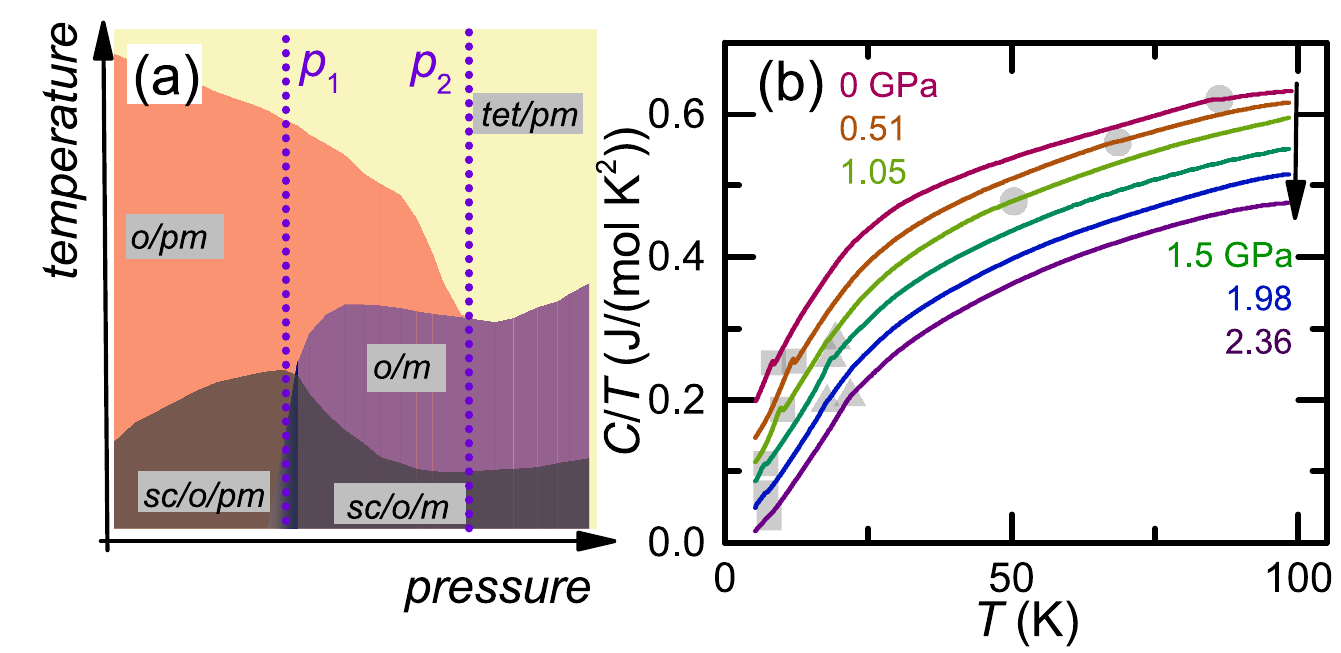}
		\caption{(a) Schematic temperature-pressure phase diagram of FeSe for hydrostatic pressures up to 2.5\,GPa. Tet (o) stands for tetragonal (orthorhombic), pm (m) for paramagnetic (magnetic) and sc for superconducting. Multiple states exist in each shaded region; for example at lowest temperature and pressure there is a superconducting, orthorhombic state with no long-range magnetic order (i.e., paramagnetic in terms of disordered moments). The purple dotted lines, labeled with $p_1$ and $p_2$, mark characteristic pressures in the phase diagram: $p_1$ marks the onset of magnetic order and $p_2$ corresponds to the pressure, at which the magnetic and orthorhombic transition merge into a single transition; (b) Selected specific heat divided by temperature, $C/T$, data plotted as a function of temperature, $T$, for different pressures in the range 0\,GPa$\,\le\,p\,\le\,2.36\,$GPa. Light grey squares, triangles and circles mark the position of the superconducting, magnetic and structural transition, respectively. Copyright American Physical Society 2019.}
		\label{fig:FeSe-schematics}
		\end{center}
		\end{figure}
	
	In more detail, the temperature-pressure phase diagram of FeSe can be described as follows (see Fig.\,\ref{fig:FeSe-schematics}\,(a) for a schematic, enlarged view on the low-pressure $p\,\lesssim\,2.5\,$GPa region of this phase diagram). At ambient pressure, FeSe undergoes a second-order nematic, tetragonal-to-orthorhombic phase transition at $T_s\,\approx\,90$\,K \cite{McQueen09} and becomes superconducting below $T_c\,\approx\,8.5\,$K \cite{Hsu08,Boehmer16c}. Upon pressurization, $T_s$ is suppressed and $T_c$ increased. Above the onset of magnetic order at $p\,\approx\,0.9\,$GPa \cite{Bendele10,Bendele12,Terashima15,Sun16,Kothapalli16}, which we will label by $p_1$ throughout the manuscript, $T_c$ is initially suppressed \cite{Miyoshi14,Kaluarachchi16,Terashima16}, and magnetic order occurs at a transition temperature $T_N$, which is well below $T_s$ ($T_N\,<\,T_s$). X-ray measurements under pressure showed that this second-order magnetic transition at $T_N$ results in an increase of the orthorhombic distortion \cite{Kothapalli16,Boehmer19}, which was interpreted as a sign of a cooperative coupling of magnetism and nematicity.  This result also supports an interpretation in favor of stripe-type magnetic order in FeSe under pressure, since the lattice symmetry is broken in the same way as for stripe-type magnetic order, e.g., in the 122 pnictides \cite{Paglione10,Dai15}. The idea of stripe-type magnetic order was further substantiated by NMR measurements under pressure \cite{Wang16c,Wiecki17}. However, a clear experimental identification of the magnetic configuration of FeSe under pressure by, e.g., neutron diffraction is still lacking likely due to the small moment of $\approx\,$0.2\,$\mu_B$/Fe, which was inferred from the $\mu$SR \cite{Bendele10,Bendele12,Khasanov17} and M\"ossbauer measurements \cite{Kothapalli16,Boehmer19}. At even higher pressures of $p\,\approx\,1.6\,$GPa (labeled by $p_2$ in the following), the magnetic and structural transition lines merge into a simultaneous, first-order transition line \cite{Kothapalli16,Boehmer19}. The magneto-structural transition line depicts a dome, centered around $5\,$GPa, with a maximum transition temperature $T_{s,N}\,\approx\,45$\,K \cite{Sun16,Boehmer19}. For $p_2\,<\,p\,\lesssim\,5\,$GPa, the superconducting $T_c$ shows an overall increase; $T_c$ becomes maximal close to the pressure at which the magneto-structural transition is suppressed ($p\,\approx\,$6\,GPa) and where FeSe remains tetragonal, but still magnetic to lowest temperatures. Finally, for even higher pressures at $\approx\,7.7\,$GPa FeSe undergoes a pressure-induced transition from a tetragonal to an orthorhombic crystal structure over a wide temperature range \cite{Boehmer19}, which likely is the reason for the loss of superconductivity at low temperatures.
	
	Whereas many investigations focused on an identification of the various ground states under pressure, many efforts have also been devoted to study the evolution of magnetic and nematic fluctuations under pressure. Despite the absence of long-range magnetic order at ambient pressure, strong magnetic fluctuations of both stripe-type and N\'{e}el-type were detected \cite{Boehmer15,Wang16,Wang16b,Kotegawa12,Rahn15,Ma17}. Upon cooling through $T_s$ at ambient pressure, the spectral weight of the magnetic fluctuations is shifted towards lower energies and towards stripe type order. Since NMR measurements revealed that magnetic fluctuations become only strongly enhanced below $T_s$, it was suggested that orbital order is the driver of nematicity at ambient pressure \cite{Baek15}. However, it was also pointed out that the presence of sizable magnetic fluctuations might be taken as an indication that magnetic frustration might play a major role for the absence of long-range order at low pressures \cite{Glasbrenner15} and thus, the absence of magnetic order might still be consistent with a spin-driven scenario. Upon pressurization, low-energy magnetic fluctuations set in below a temperature $T^\prime\,\approx\,90\,$K, which was found to be almost independent of pressure \cite{Wiecki17,Wang16c}. Based on this observation, it was argued that the coincidence of $T^\prime$ and $T_s$ at ambient pressure might be accidental and thus, might be still consistent with a spin-driven scenario. In addition, it was observed that the occurrence of the measurable low-energy magnetic fluctuations correlates with the onset of local nematicity, which is consistent with the notion of a cooperative coupling of nematicity and magnetism \cite{Wiecki17}. Overall, whereas it has been appreciated that this cooperative coupling of magnetic and structural order has led to similarities in the phase diagram of FeSe for high pressures and the archetypal 122 phase diagram, it has also been questioned whether these ideas also apply for the nematic order in the low-pressure regime. In fact, several papers recently argued on the basis of pressure-dependent measurements on FeSe that the nematic order for low and high pressures should have a distinctly different origin \cite{Khasanov18,Massat18,Boehmer19}. 
		
	Further open questions, concerning the temperature-pressure phase diagram of FeSe, relate to the interplay of superconductivity and the various normal states, i.e., the purely nematic state, the magnetic-nematic state, the magnetic-tetragonal state and the highest-pressure orthorhombic state. In particular, the simultaneous enhancement of the superconducting critical temperature $T_c$ and the magnetic-nematic transition temperature $T_{s,N}$ over most of the pressure range, in which both orders are present, is not expected in a very simplified picture of competing orders \cite{Boehmer17rev}. Instead, it was intuitively expected that a fierce competition between two orders should result in a decrease (increase) of the transition temperature of the order, which is suppressed (promoted) by the tuning parameter. As a result, this observation of a simultaneous increase of both transition temperatures has initiated ideas of a cooperative nature of superconductivity and magnetic-nematic order in FeSe under pressure \cite{Bendele10,Chen19}. In addition, a significant decrease of the magnetic hyperfine field has only been found for pressures close to the onset of magnetic order (0.9\,GPa$\,\lesssim\,p\,\lesssim\,$1.4\,GPa) \cite{Bendele10,Bendele12}, and no anomaly that could be associated with $T_c$ was observed in high-pressure NMR measurements \cite{Wang16c}. Consequently, it was speculated that superconductivity in FeSe might not even be bulk for high pressures $p\,>\,p_2$ \cite{Wang16c,Yip17}. Along these lines, it was observed that the feature of the superconducting transition in resistance measurements is significantly broader \cite{Terashima15,Chen19,Sun16}, whenever superconductivity is proposed to coexist with the magnetic-nematic order, which might be indicative of a filamentary superconducting state. The question of the bulk or filamentary nature of superconductivity over wide pressure ranges is particularly important to understand the interplay of superconductivity with different normal states. In this regard, FeSe is an important reference system, since the multitude of different normal states can be conveniently accessed by moderate hydrostatic pressures without introducing changing levels of disorder.
	
			\begin{figure*}[t]%
		\begin{center}
		\includegraphics*[width=0.9\textwidth]{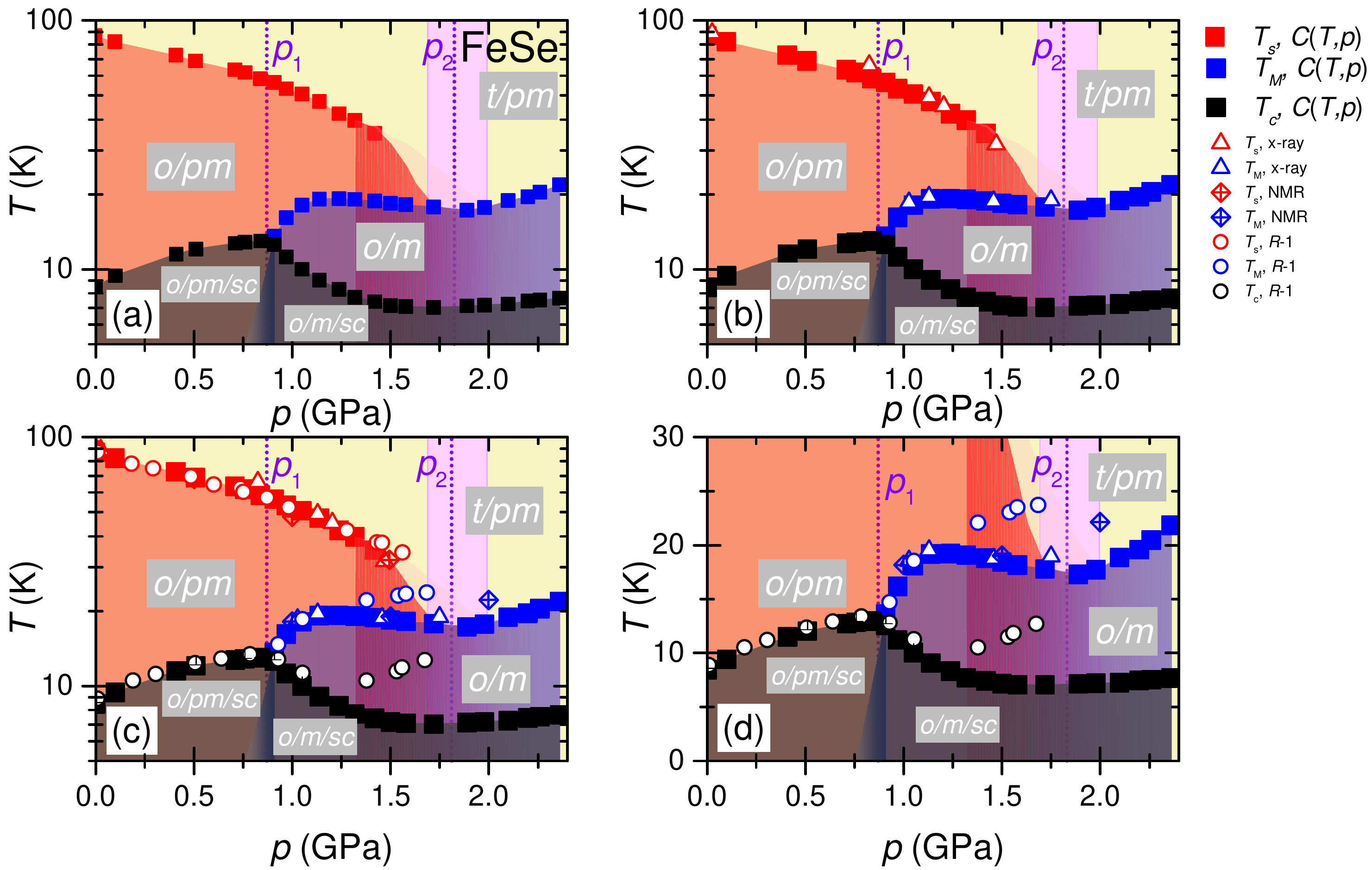}
		\caption{Comparison of the temperature-pressure phase diagram of FeSe, as constructed from specific heat measurements (solid symbols), with the phase diagrams, determined from other measurements (open symbols); (a) Temperature-pressure phase diagram, determined from specific heat measurements. Red squares denote the structural transition temperature from the tetragonal (tet) into an orthorhombic (o) state. Black squares indicate the position of the superconducting transition (sc). Blue squares mark the position of the transition from the paramagnetic (pm) to the magnetic (m) transition; (b) Comparison of the specific-heat phase diagram with the one of x-ray diffraction \cite{Kothapalli16}; (c) Comparison of phase diagram from specific heat with data from x-ray diffraction \cite{Kothapalli16}, NMR \cite{Wiecki17} and resistance \cite{Kaluarachchi16}, all taken on crystals from the same source; (d) Enlarged view on the low-temperature region of the phase diagram, presented in (c), on a linear temperature scale. Reprinted with permission from Ref. \cite{Gati19b}, Copyright American Physical Society 2019.}
		\label{fig:FeSe-simplified-phasediagram}
		\end{center}
		\end{figure*}
	
	Motivated by the search for bulk superconductivity, we performed a thermodynamic investigation of the temperature-pressure phase diagram up to a pressure of $p\,\approx\,2.36\,$GPa by utilizing measurements of the specific heat, $C$ \cite{Gati19b}. These data were obtained by the technique of $ac$ calorimetry, which was introduced in Sec.\,\ref{sec:experimentalmethods}. The optimized thermometry of our setup  \cite{Gati19a} was highly beneficial for this study, since it allows for the study of the specific heat of one single sample under pressure from low $T$, below the superconducting $T_c$, to temperatures as high as at least 100\,K, i.e., much higher than $T_s$ at ambient pressure, as can be seen in example data sets of $C/T$ vs. $T$ in Fig.\,\ref{fig:FeSe-schematics}\,(b). This therefore allows for the identification of all the salient phase transitions which are associated with the temperature-pressure phase diagram of FeSe.
		
		Now we will discuss the two central results of our specific heat study \cite{Gati19b}. First, we focus on the question of bulk superconductivity. For all pressures studied up to 2.36\,GPa, we found a specific heat feature, that can be associated with the transition into the superconducting state. The respective transition temperatures $T_c$ are depicted in the temperature-pressure phase diagram in Fig.\,\ref{fig:FeSe-simplified-phasediagram}, together with the magnetic and structural transition temperatures $T_N$ and $T_s$, which we were also able to determine from our specific heat data. Initially, for $p\,<\,p_1$, we find an increase of $T_c$ with increasing $p$. With the onset of magnetic order at $p_1$, $T_c$ is immediately suppressed with pressure, and subsequently goes through a minimum, centered around $p_2$. Above $p_2$, $T_c$ shows a very mild increase with increasing pressure again. In the same pressure range, the merged magneto-structural transition at $T_{s,N}$ also shows a positive pressure dependence, i.e., d$T_{s,N}$/d$p\,>\,0$. Thus, our thermodynamic investigations confirm a simultaneous increase of $T_c$ and $T_{s,N}$ with $p$. We want to note, though, that this result is \textit{per se} not inconsistent with the notion of competing orders, even if it is unusual. 
		
		A closer look on earlier theoretical models of competing spin-density wave and superconducting order in itinerant systems \cite{Machida81} shows that that an increase of $T_N$ with $p$ should lead to decrease in $T_c/T_N$ rather than a decrease of $T_c$ itself with $p$. Stated differently, if the magnetic order is promoted with $p$, then the superconducting order is effectively suppressed as long as d$T_c$/d$p\,<\,$d$T_{s,N}$/d$p$. That this condition is indeed satisfied for the thermodynamic $T_c(p)$ and $T_{s,N}(p)$ can be seen with bare eyes when looking on the phase diagrams in Fig.\,\ref{fig:FeSe-simplified-phasediagram}\,(a) with a logarithmic temperature scale and even more clearly in Fig.\,\ref{fig:FeSe-simplified-phasediagram}\,(d) with a linear temperature scale (solid symbols). Thus, our thermodynamic phase diagram data is fully consistent with the notion of competing superconducting and magnetic-nematic order. This result is further supported by the evolution of the superconducting jump size in the specific heat as a function of pressure (\cite{Gati19b}, not shown). Whereas the superconducting jump size is increased with increasing pressure for $p\,<\,p_1$, it becomes suppressed with increasing $p$ soon after the onset of magnetic order at $p_1$ and continues to be suppressed across $p_2$. Given that the superconducting jump size in specific heat measurements, in a simple BCS picture, measures the amount of superconducting condensation energy, the strong reduction of the jump size with the onset of magnetic order is fully compatible with a competition of magnetic order and superconductivity, leading to either a microscopic coexistence or a macroscopic phase segregation. Overall, the observation of a finite specific heat jump for any pressure strongly suggests that superconductivity is bulk across the entire pressure range, in particular also in the pressure range, in which FeSe also shows a competing magnetic-nematic order.
		
		The second result, which we obtained from a study of the thermodynamic phase diagram, is inferred from a detailed comparison to previous literature results on the temperature-pressure phase diagram, which were constructed from a variety of techniques (see Fig.\,\ref{fig:FeSe-simplified-phasediagram} (b)-(d)). Specifically, in these plots, we compare the specific heat phase diagram (solid symbols) with the ones inferred from measurements of x-ray diffraction \cite{Kothapalli16} (b), NMR \cite{Wiecki17} and resistance \cite{Kaluarachchi16} (c) (each shown by open symbols), which were all taken on single crystals from the same source and mostly in a very similar pressure environment (only the x-ray diffraction data was taken in a He pressure environment). In Ref. \cite{Gati19b}, we also included a comparison to measurements of the dc magnetization \cite{Miyoshi14} and $\mu$SR measurements \cite{Bendele10,Bendele12,Khasanov18}, which are also available in literature, but which were taken on samples of a different source. Our main observation is that, whereas the superconducting and structural transition temperatures $T_c$ and $T_s$ show a very good agreement for $p\,<\,p_1$, $T_c$ and $T_N$ inferred from the different techniques show strong discrepancies for $p\,>\,p_1$. In an attempt to reconcile these observations, we suggested in Ref. \cite{Gati19b} that our results indicate wide temperature ranges of fluctuation magnetic and superconducting orders, i.e., non-long range and non-static magnetism and superconductivity, since specific heat measurements provide the bulk and static transition temperatures. We will outline this idea in the following in more detail.
		
		For the superconducting transition, the $T_c$ from specific heat is well below the temperature, for which resistance reaches zero (see Fig.\,\ref{fig:FeSe-simplified-phasediagram} (c)) and also below the onset of diamagnetism (not shown here) for $p\,>\,p_1$. In addition, we observed a sudden change in the shape of the specific heat feature right at $p_1$ from almost mean-field-like for $p\,<p_1$ to a broader, $\lambda$-like feature for $p\,>\,p_1$. Also, the superconducting transition in resistance measurements is known to become significantly broader with the onset of long-range magnetic-nematic order, and becomes sharp again for very high pressures around 6 GPa, where magnetic-nematic order is absent \cite{Terashima15,Chen19,Sun16}. Taken together, these observations suggest an intrinsic change of the superconducting properties when entering the magnetic-nematic state, which is present for $p_1\,\le\,p\,\lesssim\,6\,$GPa. Two possible scenarios might account for the experimental observations. In scenario one, the broader specific heat feature at $T_c$ for $p\,\,\ge\,p_1$ can be interpreted as a signature of superconducting fluctuations becoming of importance in a wider temperature range above $T_c$. The onset of diamagnetism at $T\,>\,T_c$ would also be fully consistent with this picture. If this was the case, then the observed changes in the specific heat across $p_1$ are likely related to changes of the Fermi surface with the onset of magnetic order, which thus place FeSe even deeper into the BCS-BEC crossover regime \cite{Yang17}. The proximity of FeSe to the crossover from BCS to BEC superconductivity was already suggested from ambient-pressure studies of FeSe \cite{Kasahara14,Kasahara16,Watashige17,Rinott17,Hanaguri19}, owing to its small Fermi energy, which is comparable to the superconducting gap size. The alternative, second scenario invokes electronic inhomogeneity, which would give rise to static short-range orders (i.e., a spatially fluctuating state). We want to stress though, that the inhomogeneity then must be intrinsically induced by the occurence of magnetism, since no disceprancy in the superconducting $T_c$ was found for $p\,<\,p_1$. In fact, recently, such a scenario was discussed for a charge analogue of the magnetic-nematic state \cite{Yu19}. As a result, it was argued that non-bulk superconductivity might preferably form in the proximity of domain walls, that are created by magnetism and pinned by the presence of weak disorder, inevitable in any real crystal. This scenario has been coined with the term of ''fragile superconductivity'' \cite{Yu19}.
		
		Now we turn to a similar discussion of the inferred magnetic transition temperatures. Overall, we find that the magnetic transition temperatures, determined from specific heat, are at the lower bound of the ones reported in literature. To illustrate this, we first contrast our specific heat phase diagram with the phase diagram from x-ray diffraction measurements \cite{Kothapalli16} (see Fig.\,\ref{fig:FeSe-simplified-phasediagram}\,(b)), which measure an increase in orthorhombicity in response to magnetism. Since x-ray diffraction measurements therefore probe - similar to thermal expansion measurements - the change of the bulk, average lattice parameters and not local structural deformations, similar transition temperatures $T_N$ should be inferred from specific heat and x-ray diffraction. That this is indeed the case, is demonstrated in Fig.\,\ref{fig:FeSe-simplified-phasediagram}\,(b). In contrast, transition temperatures, inferred from resistance \cite{Terashima15,Kaluarachchi16} and NMR measurements \cite{Wiecki17,Wang16c} (see Fig.\,\ref{fig:FeSe-simplified-phasediagram}\,(c)) and $\mu$SR measurements \cite{Bendele10,Bendele12,Khasanov18} (not shown) all give distinctly higher transition temperatures. In fact, there appears to be a correlation of the inferred transition temperature and the time scale on which the respective techniques probe magnetism. $\mu$SR, which is sensitive to the magnetism on the fastest time scale among the techniques investigated, gives the highest transition temperatures $T_N$, followed by NMR and subsequently specific heat. This correlation therefore naturally suggests that there is a wide range of temporal fluctuating magnetic order, preceding the formation of long-range magnetic order.
		
		Overall, our finding of wide ranges of fluctuating order in the presence of strongly competing electronic orders resembles close similarities to the phase diagram of underdoped cuprates \cite{Keimer15}. For the latter material class, it is by now appreciated that charge order competes with superconductivity. In addition, wide temperature ranges of fluctuating order were reported in this underdoped regime of the cuprate phase diagram. Based on this so far purely phenomenological analogy, we assign our findings in FeSe under pressure to effects of the competition of magnetic-nematic order and superconductivity. As a result, FeSe might turn out to be an important reference system for the study of effects resulting from the competition of superconductivity and other types of electronic orders. This view is initiated by the fact that the pressure tunability of FeSe from a non-magnetic nematic to a magnetic-nematic ground state in the presence of superconductivity allowed us to correlate the onset of fluctuating superconducting and magnetic orders with the presence of the competing magnetic order without introducing additional disorder. In this sense, the temperature-pressure phase diagram of FeSe continues to offer important new insights into the phase interplay in high-temperature superconductors and is certainly worth further investigations in the future. Despite the potential surprises, which might be unraveled in future studies, one of the central goals should be an unequivocal determination of the magnetic order by neutron scattering. This result will be of particular importance for an in-depth discussion of the origin of the extended temperature ranges of fluctuating in FeSe under pressure.
		
		In addition, open and timely questions about the interplay of nematicity, superconductivity and magnetism remain for the sulfur-substituted variants FeSe$_{1-x}$S$_x$. This series received attention, as the combination of the chemical pressure, induced by isoelectronic S substitution, with physical hydrostatic pressure has provided the opportunity to separate the nematic and the magnetic order on the hydrostatic pressure axis \cite{Matsuura17,Xiang17,Yip17}. As a result, a nematic quantum critical point, which is decoupled from long-range magnetic order, was found and, correspondingly, studies of its critical properties and the role of nematic fluctuations for superconductivity were possible. For example, from detailed resistance studies \cite{Reiss20} around  the nematic critical point at $p_c\,\approx\,0.6\,$GPa in FeSe$_{0.89}$S$_{0.11}$ , it was inferred that there are changes of the effective mass across $p_c$, but no divergent behavior was observed. Thus, it was argued that the nematic fluctuations are finite, but not critical at $p_c$. In addition, no enhancement of superconducting $T_c$ was observed close to $p_c$. Based on these findings, it was proposed that the nematic quantum criticality might be quenched by a strong nematoelastic coupling to the lattice. Similar ideas were recently also brought forward by theoretical considerations \cite{Paul17}. However, the notion of a nematic quantum-critical point in the absence of magnetism was recently questioned by $\mu$SR measurements under pressure \cite{Holenstein19} on samples with the same S concentration, i.e., FeSe$_{0.89}$S$_{0.11}$.  Their main finding was that the magnetic dome spans to pressures as low as 0.6\,GPa, which is much lower than previously reported and right in the same pressure range of the previously-proposed $p_c$ for a purely nematic quantum-critical point. These issues, which are related to the temperature-pressure phase diagrams of FeSe$_{1-x}$S$_x$, should also be addressed in the future.   

	\subsection{Phase diagram of Ba(Fe$_{1-x}$Co$_x$)$_2$As$_2$ under pressure}
	\label{sec:Ba122}
	
	As outlined previously, studies of the \mbox{Ba(Fe$_{1-x}$Co$_x$)$_2$As$_2$} system have contributed significantly to our understanding of the iron-based superconductor phase diagram. In the following, we will discuss the effect of pressure as a ''clean'' tuning parameter on magnetism, nematicity and superconductivity. We will compare the temperature-pressure phase diagrams \cite{Gati19c} with those revealed in substitution studies, and highlight similarities and differences in the response of electronic order to the different tuning approaches.

	\subsubsection{Effect of hydrostatic pressure on the magnetic and nematic transition temperatures for low Co doping}
	
			\begin{figure}[t]%
		\begin{center}
		\includegraphics*[width=0.75\linewidth]{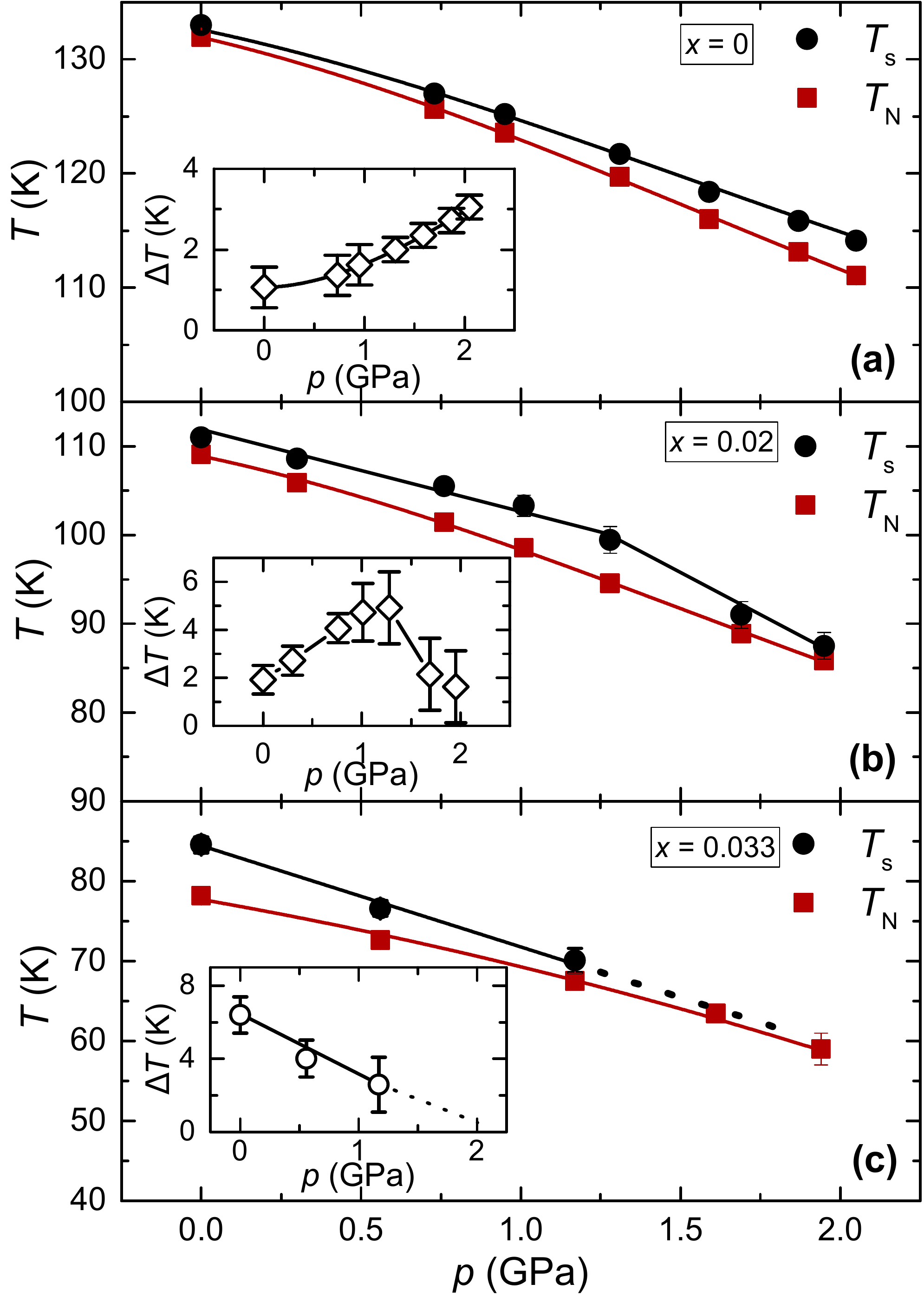}
		\caption{Evolution of antiferromagnetic transition temperature at $T_N$ (red squares) and structural transition temperature at $T_s$ (black circles) with hydrostatic pressure, $p$, in a series of Ba(Fe$_{1-x}$Co$_x$)$_2$As$_2$ with $x\,=\,$0 (a), 0.02 (b) and 0.033 (c). Solid lines are guide to the eyes. Insets: Evolution of the temperature splitting $\Delta T\,=\,T_s - T_N$ as a function of $p$ for each sample. Reprinted with permission from Ref. \cite{Gati19c}, Copyright American Physical Society 2019.}
		\label{fig:underdoped-Ba122-phasediagram}
		\end{center}
		\end{figure}
	
	In light of the observations of (i) a purely nematic state that is separated from any static and long-range magnetism in FeSe at ambient pressure \cite{McQueen09,Boehmer17rev} and (ii) a wide range of $x$ values in Ba(Fe$_{1-x}$Co$_x$)$_2$As$_2$, for which $T_s\,>\,T_N$ \cite{Canfield09,Canfield10,Kim11,Rotundu11}, as well as a limited range of $x$ for which $T_N$ is suppressed to zero but there is still a finite $T_s$ value, it is important to understand which parameter(s) controls the extent of non-magnetic, nematic order in the phase diagrams of iron-based superconductors \cite{Fernandes14}. This point has been investigated intensively by using chemical substitution as a tuning parameter in the 122 family. Whereas electron doping, as mentioned above, results in a splitting of $T_s$ and $T_N$ \cite{Canfield10}, no splitting was found in the case of hole doping or isoelectronic substitution and the magneto-structural phase transition remains a simultaneous first-order transition ($T_s\,=\,T_N$) \cite{Thaler10,Avci12}. These experimental tendencies were consistently explained by theoretical calculations, which were performed on a microscopic itinerant spin-nematic model \cite{Fernandes14,Fernandes12}. As a result of these calculations, which were based on a simplified, two-dimensional ansatz, it was proposed that the phase diagram of iron-based superconductors is controlled by a single parameter $\alpha$, which is mainly dependent on band structure parameters. Specifically, $\alpha$ was found to depend on the chemical potential $\mu$ and thus the band filling, as well as the ellipticity $\delta m$ of the electron pockets, which is a parameter that is related to the nesting of the Fermi surface. However, from an experimental point of view, it has also been pointed out that disorder might play an important role in the separation of $T_s$ and $T_N$ \cite{Nie14,Rotundu10}. It is thus of great importance to investigate the response of $T_s$ and $T_N$ to a ''clean'' tuning parameter (i.e., hydrostatic pressure), which does not involve changing levels of disorder. Whereas the effect of pressure \cite{Canfield10,Colombier09,Fukazawa08,Yamazaki10,Matsubayashi09,Ishikawa09} is known to suppress $T_s$ and $T_N$ and to induce superconductivity, the finer, more quantitative details of the phase diagram under pressure in terms of $\Delta T\,=\,T_s-T_N$ were not elucidated \cite{Wu13,Ikeda18}.
	
			\begin{figure*}[t]%
		\begin{center}
		\includegraphics*[width=0.8\textwidth]{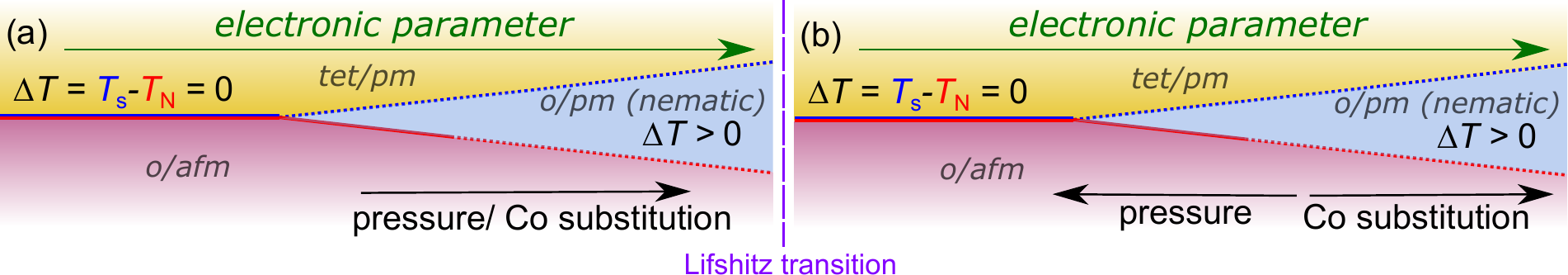}
		\caption{Schematic representation of the evolution of the temperature splitting between structural and magnetic transition, $\Delta T\,=\,T_s-T_N$, for Ba(Fe$_{1-x}$Co$_x$)$_2$As$_2$ as a function of an electronic parameter, which tunes the respective transitions from a simultaneous first-order to separated second-order transitions (indicated by the green arrow). Black arrows indicate the response of the system to Co substitution, $x$, and pressure $p$, before (a) and after (b) the system undergoes a change of Fermi surface topology. Solid red and blue lines mark the simultaneous first-order magneto-structural transition, whereas dotted red and blue lines mark the second-order magnetic and structural transition, respectively. Reprinted with permission from Ref. \cite{Gati19c}, Copyright American Physical Society 2019.}
		\label{fig:underdoped-Ba122-universal-phasediagram}
		\end{center}
		\end{figure*}
		
		With the aim to study the effect of pressure on the splitting of $T_s$ and $T_N$, we investigated the specific heat up to 2\,GPa on a series of Ba(Fe$_{1-x}$Co$_x$)$_2$As$_2$ samples with $x\,=\,0, 0.02$ and 0.033 \cite{Gati19c}. The choice for specific heat measurements was motivated by its previous success in the determination of the two transition temperatures in the series Ba(Fe$_{1-x}$Co$_x$)$_2$As$_2$ at ambient pressure \cite{Ni08,Kim11,Chu09,Rotundu10}, even though both transitions are often found to be very close in temperature. In addition, the recently optimized thermometry of our specific heat setup \cite{Gati19a} made it very promising to explore the specific heat features, associated with magnetic and structural transitions, despite the fact that the phase transitions occur at relatively high temperatures (up to 132\,K). The temperature-pressure phase diagrams, which we inferred from this study \cite{Gati19c}, are shown in Figs.\,\ref{fig:underdoped-Ba122-phasediagram}\,(a)-(c). For all three compounds, we find that overall both $T_s$ and $T_N$ are suppressed by pressure. On a gross level, this is therefore consistent with the general picture of the phase diagram of 122-type iron-based superconductors \cite{Canfield10,Colombier09,Fukazawa08,Yamazaki10,Matsubayashi09,Ishikawa09}, in which the application of pressure leads to a suppression of $T_s$ and $T_N$. However, on a finer level, the evolution of the splitting $\Delta T\,=\,T_s-T_N$ (shown in the insets of Figs.\,\ref{fig:underdoped-Ba122-phasediagram}\,(a)-(c)) shows a more complicated behavior. For the undoped parent compound, the application of pressure results in a monotonic increase of $\Delta T$ from $\,\approx\,1$K at ambient pressure up to $3.1\,$K at 2\,GPa. This behavior of $\Delta T$ can also be mapped quantitatively on the phase diagram as a function of Co substitution by using linear conversion factor between pressure and Co substitution $x$ (see \cite{Gati19c}). For the sample with $x\,=\,0.02$, we initially observed an increase of the splitting with pressure as well. However, above $p_c\,\approx\,1.3\,$GPa, $\Delta T$ suddenly is reduced with a further increase of pressure. Last, for $x\,=\,0.033$, pressure initially results in a decrease of $\Delta T$ until the two transitions likely merge around $1.5\,$GPa. Interestingly, a change of the Fermi surface topology as a function of doping in the range $0.02\,\le\,x\,\le\,0.025$ has been reported in several earlier reports \cite{Liu10,Mun09,Hodovanets13}. It is thus tempting to assume that the sign change of the initial slope of the $\Delta T(p)$ behavior as a function of doping $x$ is related to this change of Fermi surface topology. To support this hypothesis, we complemented the pressure-dependent specific heat data by measurements of the Hall effect under pressure \cite{Gati19c} on the sample with $x\,=\,0.02$, for which a sign change of d$(\Delta T)/$d$p$ can readily be induced by crossing $p_c$. Indeed, these Hall coefficient data showed an anomalous behavior right around $p_c$. Thus, this result strongly suggests that the sharp kink in the $\Delta T(p)$ behavior for the $x\,=\,0.02$ is related to a change of Fermi surface topology.
		
		Taken together, the result of our pressure study on selected underdoped samples of Ba(Fe$_{1-x}$Co$_x$)$_2$As$_2$ is summarized in Fig.\,\ref{fig:underdoped-Ba122-universal-phasediagram} in a schematic diagram \cite{Gati19c}, which compares the effect of pressure with the one of Co substitution $x$.  Initially, starting from the parent compound, pressure and Co substitution act very similarly in terms of the splitting (see Fig.\,\ref{fig:underdoped-Ba122-universal-phasediagram} (a)). The breakdown of this analogy, which is associated with a distinctly different response of the splitting to Co substitution vs. pressure (see Fig.\,\ref{fig:underdoped-Ba122-universal-phasediagram} (b)), can be correlated with a change of the Fermi surface topology, which can either be induced by pressure or Co substitution. Given that no structural changes have been reported for BaFe$_2$As$_2$ as a function of pressure or doping, this result strongly suggests that the evolution of $\Delta T$ is governed by some parameter of electronic origin. In this picture, a change of the Fermi surface topology results in a non-monotonic evolution of $\Delta T$ as a function of experimental tuning parameters, such as pressure. This interpretation of our experimental results is fully consistent with the model calculations of Refs. \cite{Fernandes14,Fernandes12}. A future goal is to use this experimental benchmark, which provides clear critical pressures and concentrations, for a refined microscopic modeling of the behavior of magnetism and nematicity in the phase diagram of iron-based superconductors. To this end, band structures, which were obtained from detailed density-functional theory calculations, across the change of Fermi surface topology might be used as a more realistic starting point for a microscopic model. However, it has to be said that a first attempt to identify the change of Fermi surface topology in DFT calculations has turned out to be difficult \cite{Borisov20}, likely due to the importance of correlations for the detailed band structure calculations.
	
	\subsubsection{Effect of hydrostatic pressure on superconductivity beyond optimal doping}
	
Whereas the previous section focused on the comparison of the effect of doping and pressure on the nematic and magnetic transition in underdoped Ba(Fe$_{1-x}$Co$_x$)$_2$As$_2$ compounds, we want to discuss here the effect of fine-tuned hydrostatic pressure on superconducting samples of Ba(Fe$_{1-x}$Co$_x$)$_2$As$_2$ with $x\,=\,0.075$, 0.093 and 0.112, which are all located in the overdoped regime of the phase diagram ($x\,>\,x_{opt}\,\approx\,0.06 - 0.07$) \cite{Canfield10}. To obtain thermodynamic information about the temperature-pressure phase diagram, we performed magnetization measurements up to $\approx\,$1\,GPa in a piston-pressure cell on samples of the three concentrations listed above. Given our results of a non-linear change of $T_c$ with $p$ up to $\approx\,1.3\,$GPa for the samples with $x\,=\,$0.075 and 0.093, which we will discuss below, we additionally measured the sample with $x\,=\,0.075$ up to $\approx\,$2\,GPa via specific heat measurements in piston-pressure cells in order to check for a possible sign change of the slope d$T_c/$d$p$ for this particular sample for higher pressures. The details of the experimental techniques were introduced in detail in Sec.\,\ref{sec:experimentalmethods}.

\begin{figure}[t]%
		\begin{center}
		\includegraphics*[width=\linewidth]{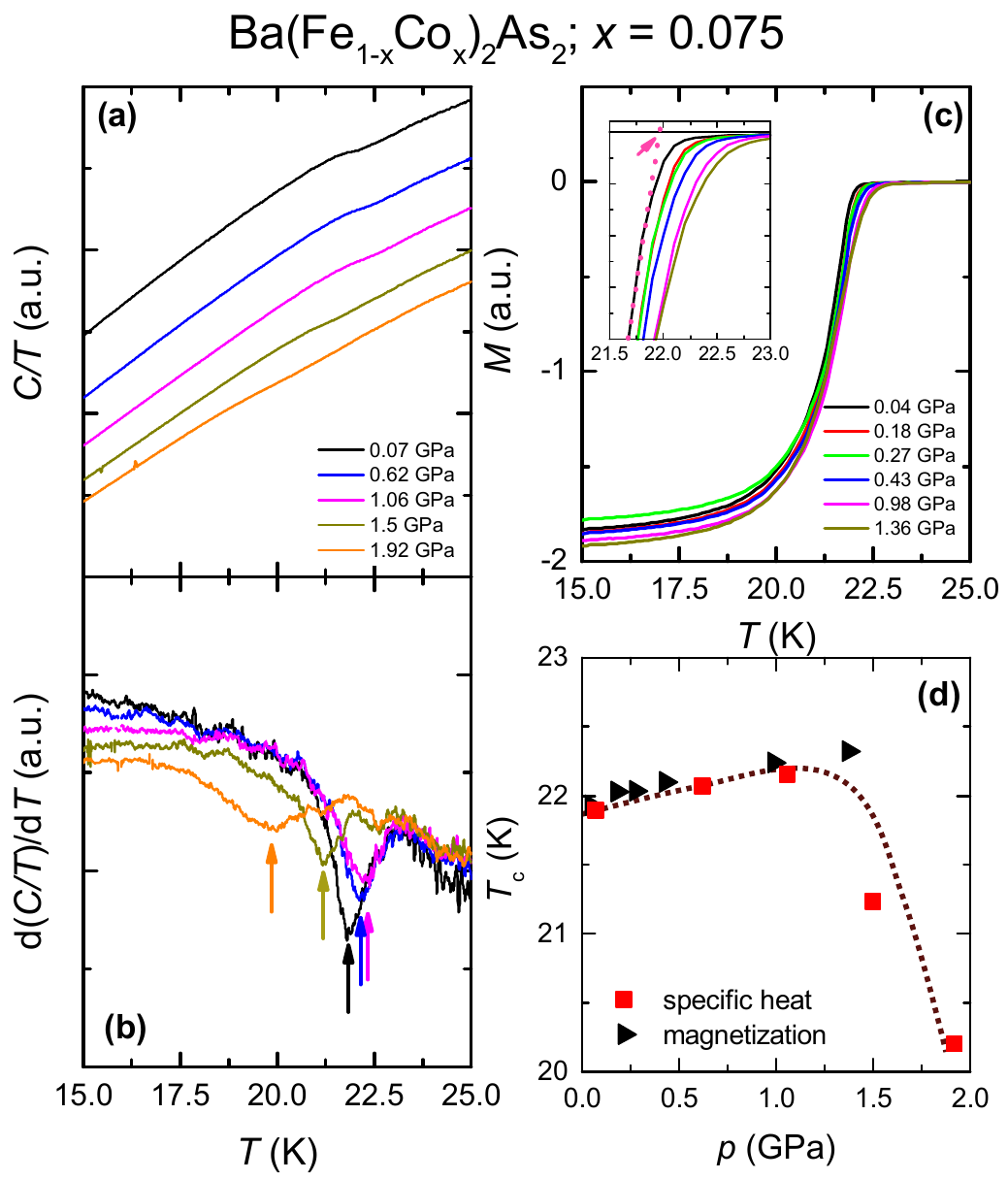}
		\caption{Temperature-pressure phase diagram of Ba(Fe$_{1-x}$Co$_x$)$_2$As$_2$ with $x\,=\,0.075$. (a) Data of the specific heat divided by temperature, $C/T$, and (b) its temperature derivative, d$(C/T)/$d$T$, for hydrostatic pressures in the range 0.07\,GPa$\,\,\le\,p\,\le\,$1.92 GPa. Arrows in (b) indicate the position of a minimum in d$(C/T)/$d$T$, which is used as a criterion to determine the superconducting transition temperature $T_c$; (c) Magnetization, $M$, vs. $T$ data up to 1.36\,GPa upon increasing pressure. Inset: $M(T)$ data on enlarged scales around $T_c$. The dashed-dotted line is used to visualize the criterion to determine $T_c$ from the $M$ data; (d) Temperature-pressure phase diagram for a sample of Ba(Fe$_{1-x}$Co$_x$)$_2$As$_2$ with $x\,=\,0.075$. Red squares (black triangles) mark the superconducting transition temperature $T_c$, determined from specific heat measurements (magnetization measurements). The dotted brown line is a guide to the eyes.}
		\label{fig:overdoped-75-Ba122-data}
		\end{center}
		\end{figure}

The specific heat, $C(T)$, and magnetization, $M(T)$ data for a Ba(Fe$_{1-x}$Co$_x$)$_2$As$_2$ sample with $x\,=\,0.075$ for different applied pressures are shown in Fig.\,\ref{fig:overdoped-75-Ba122-data}. To determine the change of $T_c$ with $p$, we used the following criteria. For specific heat measurements, we determined the position of the minimum in d$(C/T)$/d$T$ and assigned it to $T_c$ (see Fig.\,\ref{fig:overdoped-75-Ba122-data} (b)). In the case of magnetization measurements, we used the intersection of two straight lines, which represent extrapolations from below and above the onset of diamagnetism, to determine $T_c$. The resulting temperature-pressure phase diagram is shown in Fig.\,\ref{fig:overdoped-75-Ba122-data} (d). For low pressures, both data sets show consistently a very moderate increase of $T_c$ with $p$ up through the highest pressure magnetization data point at 1.3 GPa with an average slope d$T_c$/d$p\,\approx\,$+0.2\,K/GPa over the whole pressure range. Since the magnetization setup is limited to pressures below $\approx\,$1.3\,GPa, we can only rely on specific heat data for higher pressures. Surprisingly, the specific heat data suggest that beyond 1.3\,GPa $T_c$ decreases rapidly with pressure. In summary, $T_c(p)$ thus exhibits a non-monotonic variation, with a maximum around $p_{max}\,\approx\,$1.3\,GPa and a pronounced asymmetry between the low-pressure ($p\,<\,p_{max}$) and high-pressure ($p\,>\,p_{max}$) behavior. 
	
			\begin{figure}[t]%
		\begin{center}
		\includegraphics*[width=0.7\linewidth]{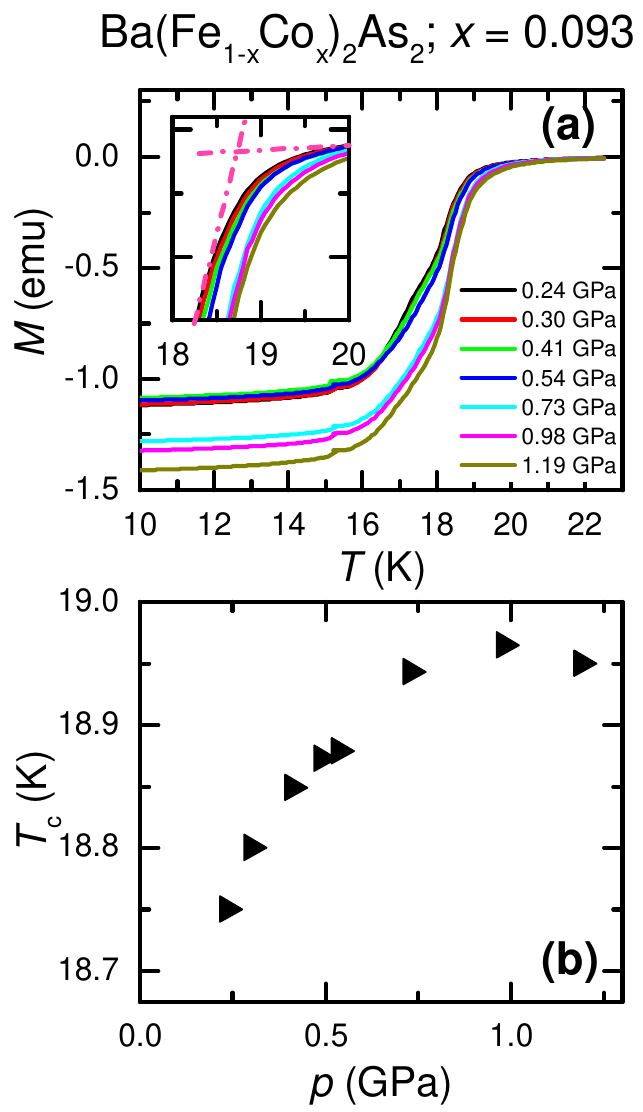}
		\caption{Temperature-pressure phase diagram of Ba(Fe$_{1-x}$Co$_x$)$_2$As$_2$ with $x\,=\,0.093$. (a) Magnetization, $M$, vs. $T$ data up to 1.19\,GPa upon increasing pressure. Inset: $M(T)$ data on enlarged scales around $T_c$. Dashed-dotted lines are used to visualize the criterion to determine $T_c$ from the $M$ data; (b) Temperature-pressure phase diagram, determined from magnetization measurements.}
		\label{fig:overdoped-9-Ba122-data}
		\end{center}
		\end{figure}
		
\begin{figure}[t]%
		\begin{center}
		\includegraphics*[width=0.7\linewidth]{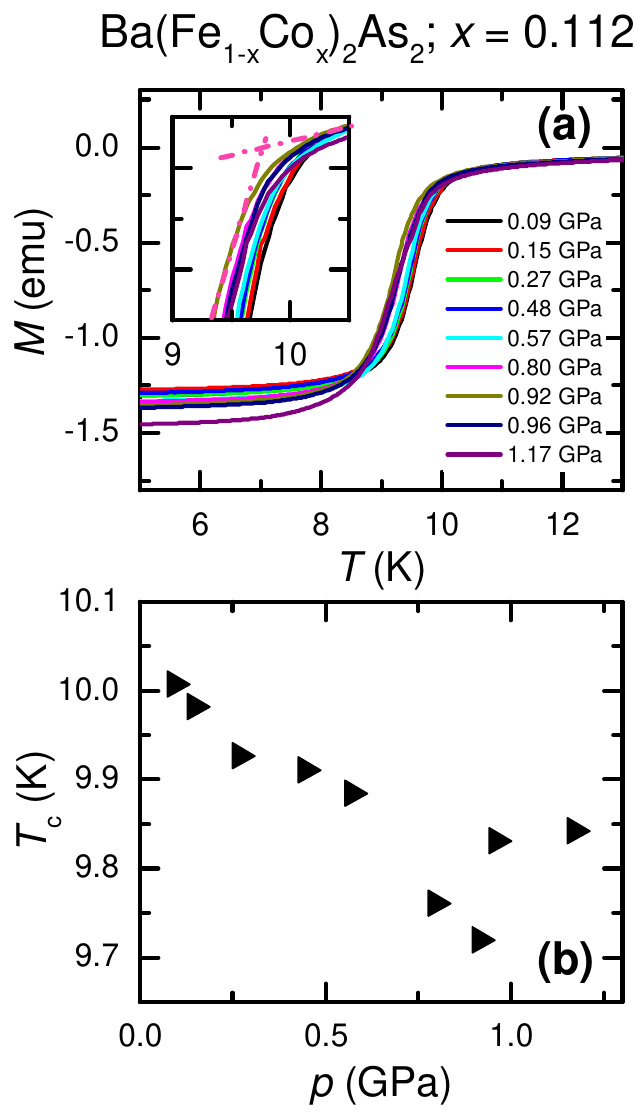}
		\caption{Temperature-pressure phase diagram of Ba(Fe$_{1-x}$Co$_x$)$_2$As$_2$ with $x\,=\,0.112$. (a) Magnetization, $M$, vs. $T$ data up to 1.17\,GPa upon increasing pressure. Inset: $M(T)$ data on enlarged scales around $T_c$. Dashed-dotted lines are used to visualize the criterion to determine $T_c$ from the $M$ data; (b) Temperature-pressure phase diagram, determined from magnetization measurements. The somewhat larger scattering in $T_c$ values can be attributed to a small change in slope of the extrapolation line below $T_c$, likely due to minor reorientation of the single crystal.}
		\label{fig:overdoped-11-Ba122-data}
		\end{center}
		\end{figure}

		Before discussing a possible interpretation of this data, we will first discuss how the two other overdoped Ba(Fe$_{1-x}$Co$_x$)$_2$As$_2$ systems respond to pressure. Figure \ref{fig:overdoped-9-Ba122-data} shows magnetization data, $M(T)$, and the temperature-pressure phase diagram of a sample with $x\,=\,0.093$. Similar to the behavior of the sample with $x\,=\,0.075$, $T_c$ initially increases with increasing pressure, until it flattens around 1\,GPa and starts to gradually decrease. In contrast, for the sample with $x\,=\,0.112$, for which the data and the phase diagram are shown in Fig.\,\ref{fig:overdoped-11-Ba122-data}, $T_c$ is suppressed from the ambient pressure value with increasing pressure over the whole pressure range investigated.
		
		The results presented here are remarkable since they reveal an initial increase and suggest a maximum of $T_c(p)$ for the two samples with $x$ values which are closer to optimal doping ($x\,=\,0.075$ and 0.093). Our results are consistent with previous reports on samples from the same source \cite{Colombier10}, and somewhat consistent with samples from other sources \cite{Arsenijevic11,Ahilan09,Rosa15}. Some of the latter studies reveal a constant, positive slope of $T_c(p)$ up to 2\,GPa for $x\,\le\,0.099$, whereas others show a maximum of $T_c(p)$ with a strong decrease of $T_c$ for higher pressures up to 2\,GPa, in accordance with our data. The origin of these discrepancies for higher $p$ for the different studies is unclear at present. Irrespective of the detailed $T_c(p)$ behavior for high pressures, all measurements on overdoped samples for $x\,\lesssim\,0.1$ are consistent in the sense that doping and pressure have a clearly different effect on $T_c$, since they suppress and support the formation of superconductivity, respectively. Only for the sample with $x\,=\,0.112$, the application of pressure has the same effect on $T_c$ as an increase in Co substitution. In fact, the pressure response of $T_c$ was already discussed intensively in high-temperature cuprate superconductors \cite{Lorenz05,Neumeier93}, with a strong focus on the questions (i) why the sign of d$T_c$/d$p$ is positive for many overdoped members, and (ii) why $T_c(p)$ shows a non-linear behavior. Based on these investigations, it was pointed out that pressure cannot be simply mapped onto a change of carrier density, i.e., d$T_c$/d$p$\,=\,d$T_c^i$/d$p$+d$T_c$/d$n$ d$n$/d$p$ \cite{Neumeier93}, with $n$ the charge carrier density. In this equation, the first term d$T_c^i$/d$p$ accounts for effects that are unrelated to a charge-carrier density change, induced by pressure. Such an ansatz accounts for the observation of a positive slope of $T_c(p)$ or local maxima and minima in $T_c(p)$. However, the microscopic mechanism behind the term d$T_c^i$/d$p$ are likely complex, and material-dependent. Another aspect, which might explain many of our observations on Co-doped BaFe$_2$As$_2$ and should likely be taken into account, is that a Lifshitz transition has also been reported for the doping range around $0.1\,\le\,x_c\,\le\,0.12$ \cite{Liu10,Mun09,Hodovanets13,Ni11}, which is associated with the disruption of a neck at the $\Gamma$-point. This might suggest a possible scenario in which, again, discontinuous changes of the Fermi surface might be responsible for a sign change of the initial d$T_c$/d$p$ from $x\,\le\,x_c$ to $x\,\ge\,x_c$. Along these lines, it is noteworthy that the position of the maximum in $T_c(p)$ is shifted from $\approx$\,1.3\,GPa for $x\,=\,0.075$ to $\approx$\,1.0\,GPa for $x\,=\,0.093$. If the maximum $T_c$ would be associated with an electronic Lifshitz transition, then our results would suggest that increasing $x$ pushes the Lifshitz transition to lower pressures and as a result, for high enough $x$, such as e.g., $x\,=\,0.112$, only a negative pressure dependence can be detected. We stress though that this is, as of now, only a hypothesis that could explain the non-monotonic pressure dependence of $T_c$ with different signs of the slope, d$T_c$/d$p$. Conversely, the role of disorder, induced by Co substitution, might warrant further consideration in order to provide an explanation for the different effect of Co substitution and pressure. Overall, we can summarize that the conventional wisdom of an analogy of pressure and Co doping is not fulfilled in all details for the evolution of the superconducting $T_c$ in the overdoped regime. If the nature of d$T_c^i$/d$p$ would be known, this would likely allow for the inference of information about the superconducting mechanism.

\section{Tuning of the collapsed-tetragonal transition by hydrostatic pressure and impact on electronic properties}
\label{sec:Casystems-cT}

In this section, we will discuss how pressure tuning has allowed for the discovery of the collapsed-tetragonal structural phase transition in 122 and 1144 structure classes of iron-based superconductors, which is associated with the interlayer-bonding of As-As orbitals, in representative iron-based superconductors as well as for the study of its impact on the electronic properties. In doing so, we will focus on (i) the series of Ca(Fe$_{1-x}$Co$_x$)$_2$As$_2$, owing to its extraordinary high pressure sensitivity, which makes the collapsed tetragonal (cT) phase transition readily accessible to relatively low pressure, laboratory experiments \cite{Torikachvili08,Yu09,Goldman09,Kreyssig08,Gati12} and (ii) the series of CaK(Fe$_{1-x}$Ni$_x$)$_4$As$_4$ \cite{Meier18}, in which a new type, the so-called half-collapsed tetragonal (hcT) structure was discovered \cite{Kaluarachchi17,Borisov18}, as a consequence of the layer-by-layer-segregation of alkali- and alkali-earth ions with very different radii along the $c$ axis. At the end of this section, we will shortly outline how these specific structural transitions have lead to the observation of remarkable elastic properties (''superelasticity'') \cite{Sypek17,Song19} and emphasize why these materials present a promising platform for strain engineering.

\subsection{Effect of pressure on Ca(Fe$_{1-x}$Co$_x$)$_2$As$_2$: Transition from the tetragonal to the collapsed tetragonal structure}
	\label{sec:Ca122}
	
	The parent compound CaFe$_2$As$_2$ \cite{Ni08b,Ronning08,Canfield09b} undergoes a very sharp first-order, simultaneous magnetostructural phase transition from the high-temperature tetragonal, paramagnetic to the low-temperature orthorhombic, antiferromagnetic  state at $T_{s,N}\,\approx\,170$\,K. Early pressure experiments \cite{Torikachvili08,Park08}, using a piston-pressure cell with liquid pressure medium, suggested that superconductivity can be stabilized by very moderate pressures 0.25\,GPa$\,\le\,p\,\le\,$0.5\,GPa, once the magneto-structural transition is sufficiently suppressed. In addition, for $p\,\ge\,$0.5\,GPa, a transition from the high-$T$ tetragonal to the low-temperature, non-magnetic collapsed-tetragonal (cT) phase was observed \cite{Goldman09,Kreyssig08,Torikachvili08,Yu09}, which was understood to be a result of the orbital bonding of As-$p_z$ orbitals \cite{Tomic12,Diehl14}. Later pressure measurements \cite{Yu09}, which were performed by using $^4$He as a pressure-transmitting medium, were not able to detect superconductivity, whereas they confirmed the suppression of $T_{s,N}$ as well as the occurence of the cT phase upon pressurization. To reconcile these two different observations in terms of the appearance of a superconducting dome, it was shown \cite{Canfield09,Prokes10} that, when CaFe$_2$As$_2$ is cooled through the cT phase transition, which is associated with a large and anisotropic change of the sample's dimensions, while the pressure medium is solid, a multi-crystallographic-phase state might be stabilized. The associated phase separation then might lead to a partial superconductivity in the sample, which is detected in resistance measurements \cite{Prokes10}. Correspondingly, when measurements are performed under He-gas pressure, the medium is still liquid at the cT transition and the crystal is not hindered in its expansion. Thereby the formation of a multi-crystallographic-phase state is avoided, and only a transition from afm/o to cT without any evidence for superconductivity was observed under hydrostatic pressure. As a matter of fact, CaFe$_2$As$_2$ is so sensitive to external pressures that it was even possible to tune the ground state of this system from antiferromagnetic/orthorhombic (o/afm) to non-magnetic cT by postgrowth thermal annealing and quenching of single crystals grown out of FeAs self-flux \cite{Ran11,Ran14}. Based on this finding, it was suggested that postgrowth thermal annealing mimics the effect of hydrostatic pressure \cite{Ran11,Ran12,Ran14,Canfield16}. This idea was further motivated by the observation of FeAs nanoprecipitates, that are associated with a small width of formation of the CaFe$_2$As$_2$ crystals.  The size and the spatial distribution of these particles (and their associated strain on the CaFe$_2$As$_2$ matrix) can be controlled by post-growth annealing temperature $T_{anneal}$ \cite{Ran11}. It was argued that the homogeneous distribution of the nanoprecipitates likely leads to an uniform strain field on the crystals.
	
	Bulk superconductivity was stabilized in CaFe$_2$As$_2$ by Co doping on the Fe site. From a systematic study of the three-dimensional $T$-$x$-$T_{anneal}$ phase diagram of Ca(Fe$_{1-x}$Co$_x$)$_2$As$_2$ at ambient pressure \cite{Ran12,Ran14}, it was found that superconductivity emerges for sufficiently large $x$ between the o/afm and the cT phase on the $T_{anneal}$-axis. No indications for the coexistence of the superconductivity with any of the other two phases were found. For a more careful fine-tuning across the phase diagram, we determined the phase diagram of a sample of Ca(Fe$_{1-x}$Co$_x$)$_2$As$_2$ with $x\,=\,0.028$ and $T_{anneal}\,=\,350^\circ$C under hydrostatic $^4$He gas pressure \cite{Gati12}. For this particular sample, the first-order magnetostructural transition is already suppressed to $T_{s,N}\,\approx\,50\,$K at ambient pressure, and for this particular doping level $x$, postgrowth thermal annealing with $T_{anneal}\,\ge\,400^\circ$C was sufficient to stabilize superconductivity \cite{Ran12}. This therefore identifies this compound as very promising for pressure-tuning across the many salient ground states of Ca(Fe$_{1-x}$Co$_x$)$_2$As$_2$.
	
			\begin{figure}[t]%
		\begin{center}
		\includegraphics*[width=0.9\linewidth]{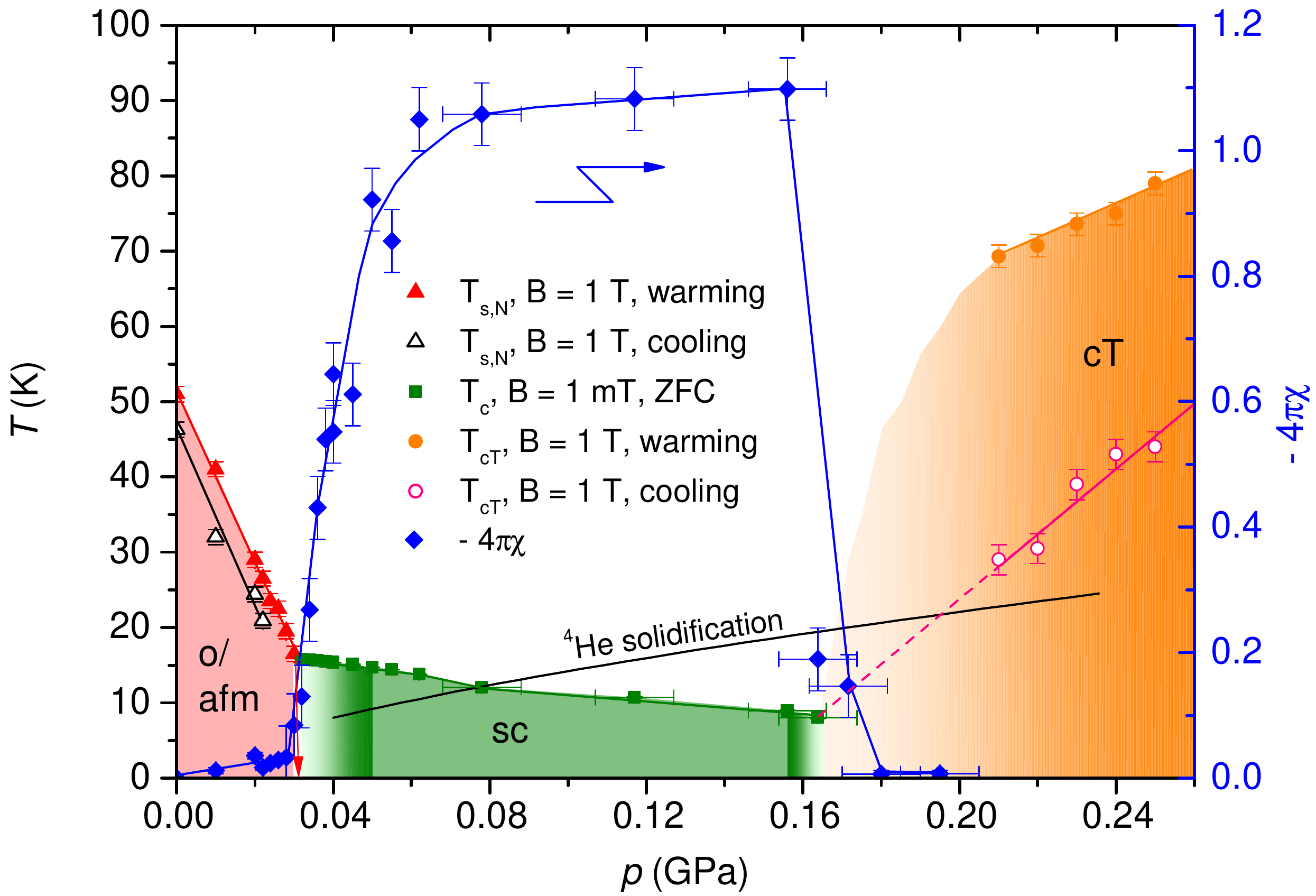}
		\caption{Temperature-pressure phase diagram of a crystal of Ca(Fe$_{1-x}$Co$_x$)$_2$As$_2$ with $x\,=\,0.028$ and $T_{anneal}\,=\,350^\circ$C (see text for a discussion of the role of annealing). The phase diagram was constructed from measurements of the magnetic susceptibility in a Helium gas-pressure cell. Filled red (open black) triangles indicate the transition from the high-temperature tetragonal, paramagnetic state to the low-temperature orthorhombic, antiferromagnetic state at $T_{s,N}$. Filled green squares correspond to the transition into the superconducting state at $T_c$, with the closed blue diamonds denoting the size of the diamagnetic shielding fraction (without correction of demagnetization effects). Filled orange (open pink) circles correspond to the transition into the full-collapsed tetragonal (cT) structure. Black line indicates the solidification line of the Helium pressure medium. Reprinted with permission from Ref. \cite{Gati12}, Copyright American Physical Society 2012.}
		\label{fig:Ca122-phasediagram}
		\end{center}
		\end{figure}
		
		The resulting temperature-pressure phase diagram for Ca(Fe$_{1-x}$Co$_x$)$_2$As$_2$ with $x\,=\,0.028$ and $T_{anneal}\,=\,350^\circ$C, which was constructed from a combination of magnetization and resistance measurements under $^4$He-gas pressure, is shown in Fig.\,\ref{fig:Ca122-phasediagram} \cite{Gati12}. As mentioned above, the use of He-gas as a pressure transmitting medium is particularly important for this system, given its high pressure sensitivity and the associated sensitivity to non-hydrostatic pressure components. Indeed, pressures of only 0.03\,GPa are sufficient to suppress the o/afm transition to zero, resulting in an extraordinary high pressure coefficient of d$T_{s,N}$/d$p\,=\,-(1100\,\pm\,50)\,$K/GPa. Over this pressure range, the o/afm transition remains first-order and no indications for a significant diamagnetic shielding volume (blue line on the right axis in Fig.\,\ref{fig:Ca122-phasediagram}), associated with a possible superconducting phase, can be found. For higher pressures, a superconducting (sc) state with essentially full shielding volume was stabilized up to $p\,\,\approx\,$0.16\,GPa, and the superconducting $T_c$ is suppressed with increasing $p$ by d$T_{c}$/d$p\,=\,-(60\,\pm\,3)\,$K/GPa. For even higher pressures $p\,\ge\,0.21\,$GPa, clear indications for a temperature-induced transition from the tetragonal to the collapsed-tetragonal structure were found, and the corresponding transition temperature $T_{cT}$ increases with increasing $p$ by d$T_{cT}$/d$p\,=\,+(420\,\pm\,70)\,$K/GPa. At the same time, no shielding volume was detected, whenever the cT transition took place upon cooling. Altogether, all salient ground states associated with iron-based superconductors (o/afm, sc and cT) can be accessed here in one single sample using modest and truly hydrostatic pressures. These studies also revealed no sign of any coexistence of superconductivity with the nearby o/afm and nonmagnetic cT phases. This observation was related to the strongly first-order character of the o/afm as well as the cT phase transition (see also \cite{Knoener16}). As a result, we argued that these results indicate that preserving fluctuations to low enough temperatures is vital for sc to form here \cite{Fernandes10}. Similar conclusions were inferred later from neutron scattering \cite{Soh13}, ARPES \cite{Dhaka14} and NMR measurements \cite{Furukawa14} across the temperature-induced cT transition, which can be accessed at ambient pressure by using postgrowth thermal annealing at a temperature $T_{anneal}$ as a tuning parameter.
		
				\begin{figure}[t]%
		\begin{center}
		\includegraphics*[width=\linewidth]{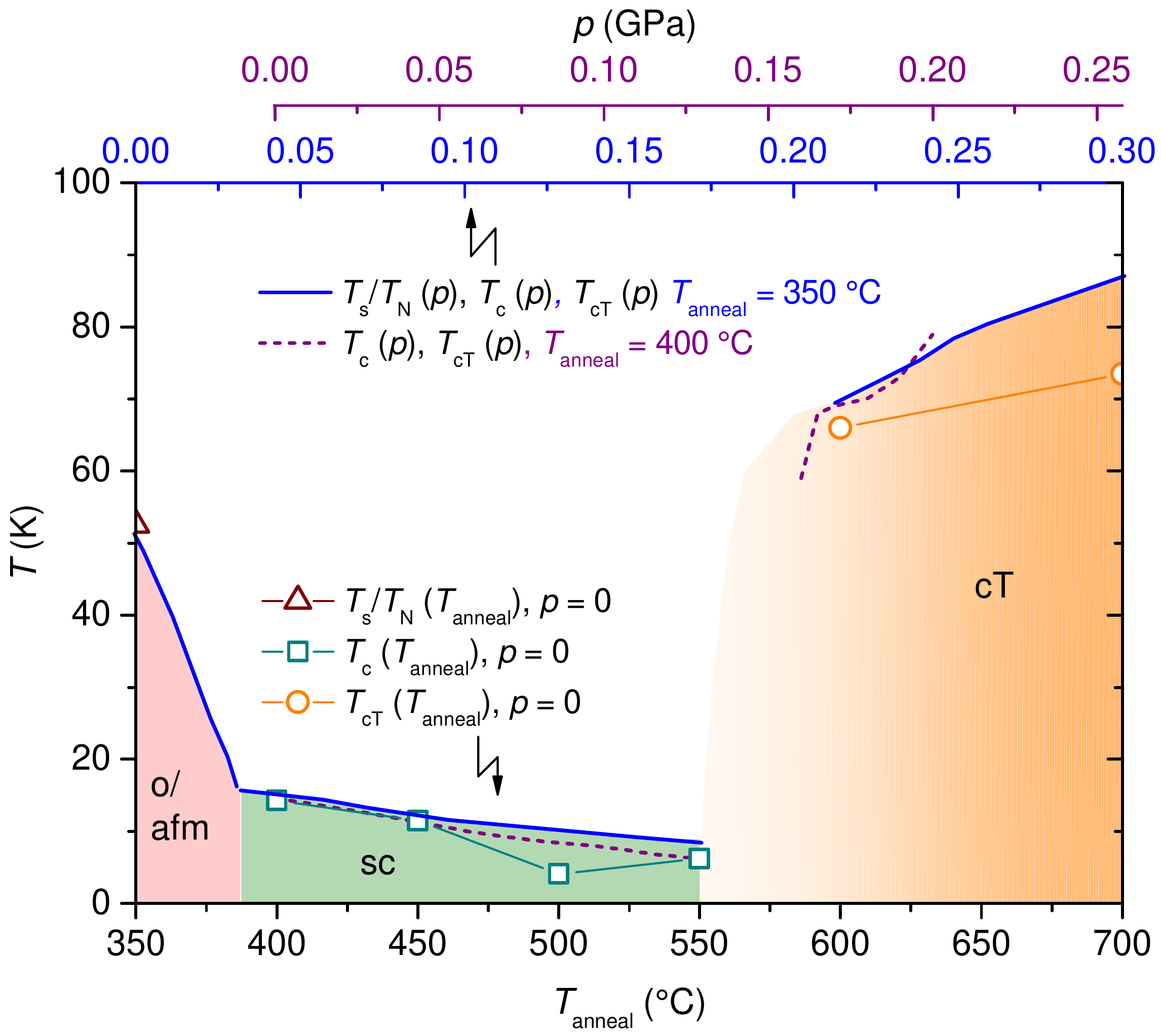}
		\caption{Composite phase diagram of single crystals of Ca(Fe$_{1-x}$Co$_x$)$_2$As$_2$ with $x\,=\,0.028$ with annealing temperatures $T_{anneal}$ and hydrostatic pressure $p$ as tuning parameters. The bottom axis is linked to phase diagram data (shown by symbols) for a sample at ambient pressure ($p\,=\,0$) with $T_{anneal}$ as a tuning parameter (taken from Refs. \cite{Ran12,Ran14}). Top blue (purple) axis is linked to phase diagram data (shown by solid (dotted) lines) for a sample with $T_{anneal}\,=\,350^\circ$C ($T_{anneal}\,=\,400^\circ$C) and $p$ as a tuning parameter (taken from Ref. \cite{Gati12}). Red area denotes the range of orthorhombic/antiferromagnetic order, green area the range of superconductivity, and orange area the range of collapsed tetragonal (cT) structure.}
		\label{fig:Ca122-unified}
		\end{center}
		\end{figure}

	To establish the correspondence of hydrostatic pressure, $p$, and postgrowth thermal annealing, $T_{anneal}$, we combine the phase diagrams as a function of both tuning parameters in a single phase diagram, which is shown in Fig.\,\ref{fig:Ca122-unified}. To this end, we compared the following three phase-diagram data sets of Ca(Fe$_{1-x}$Co$_x$)$_2$As$_2$ with $x\,=\,0.028$ with (i) $T_{anneal}$ as a tuning parameter at ambient pressure (bottom axis) \cite{Ran12}, and (ii,iii) with $p$ as a tuning parameter (top axes) \cite{Gati12} for samples with fixed $T_{anneal}\,=\,350^\circ$C (ii) and $T_{anneal}\,=\,400^\circ$C (iii). The latter sample is superconducting at ambient pressure, and modest hydrostatic pressures suppress superconductivity and induce a cT phase. On a first glance, this already suggests a very similar effect of postgrowth thermal annealing at $T_{anneal}$ and hydrostatic pressure. Even on a quantitative level, a very good agreement can be achieved by using the conversion factor $\Delta T_{anneal}\,=100\,^\circ$C$\,\approx\,\Delta p\,=\,0.0846$\,GPa. Small discrepancies in the absolute transition transition temperature occur mostly for high $T_{anneal}$ and/or high $p$ in the region of the phase diagram, where the sample undergoes the cT transition. We can only speculate that these minor differences are related to the strongly anisotropic, and large thermal expansion through the cT transition \cite{Budko13,Soh13,Ran11,Budko10}, combined with slightly different strain fields, created by hydrostatic pressure vs. the nanoprecipitates \cite{Ran11}, e.g., in the out-of-plane vs. in-plane strain values. Overall, however, the very good agreement of the different phase transition lines (o/afm, sc and cT) by using a linear scaling is in full accordance with the view that $T_{anneal}$ mimics the effect of hydrostatic pressure. This analogy also supports the view, that the pressure-induced superconductivity, which was observed in a sample with $x\,=\,0.028$ and $T_{anneal}\,=\,350^\circ$C (see Fig.\,\ref{fig:Ca122-phasediagram}) via the detection of a full shielding volume in magnetization measurements, is indeed of bulk character.

		\begin{figure}[t]%
		\begin{center}
		\includegraphics*[width=0.9\linewidth]{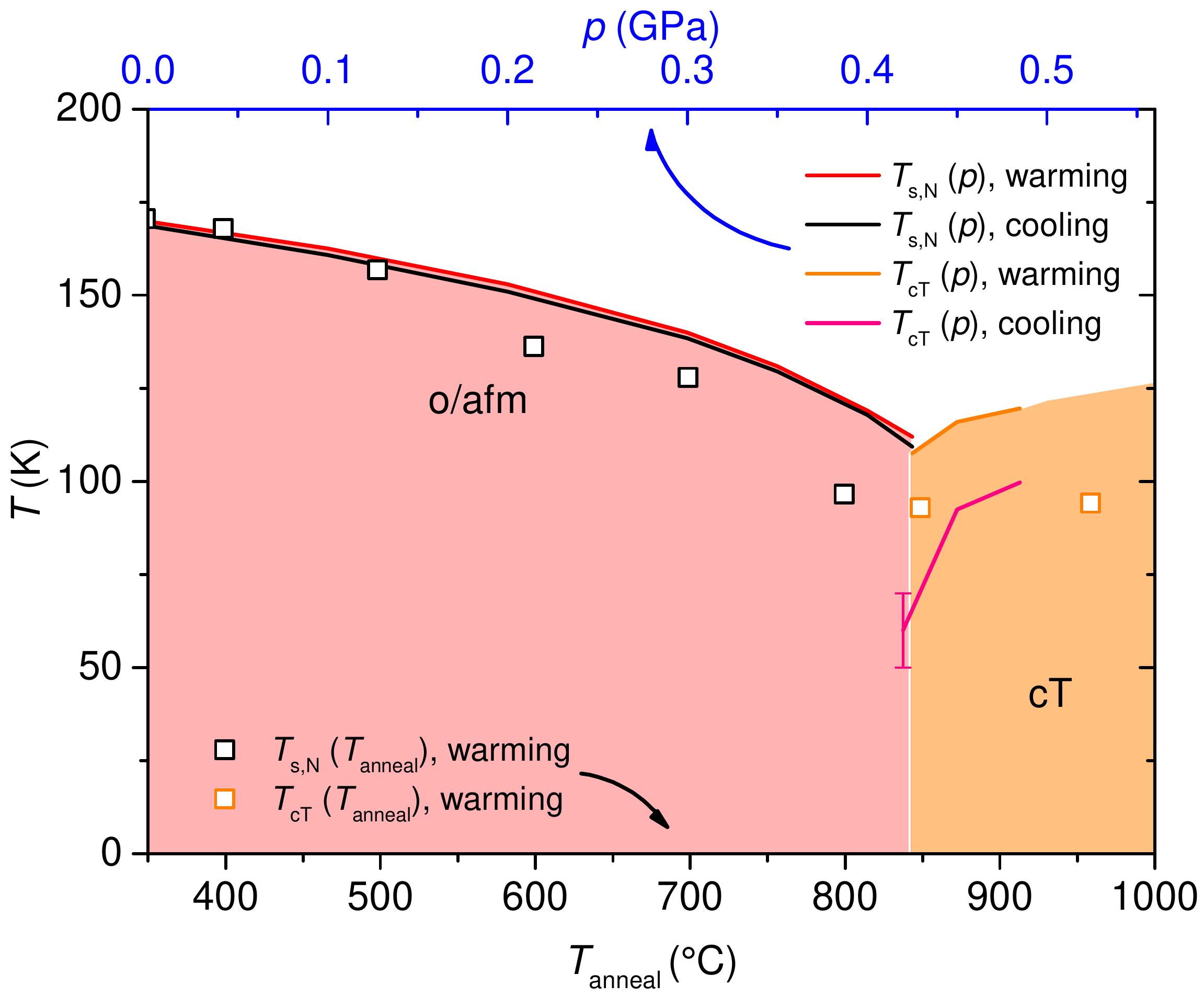}
		\caption{Composite phase diagram of single crystals of CaFe$_2$As$_2$ with annealing temperatures $T_{anneal}$ and hydrostatic pressure $p$ as tuning parameters. The bottom axis is linked to phase diagram data (shown by symbols) for a sample at ambient pressure ($p\,=\,0$) with $T_{anneal}$ as a tuning parameter (taken from Refs. \cite{Ran12,Ran14}). Top blue axis is linked to phase diagram data (shown by solid lines) for a sample with $T_{anneal}\,=\,350^\circ$C and $p$ as a tuning parameter. Red area denotes the range of orthorhombic/antiferromagnetic order, and orange area the range of collapsed tetragonal (cT) structure. For each transition, the phase transition temperatures upon warming and cooling are shown to demonstrate their first-order nature.}
		\label{fig:Ca122-unified-undoped}
		\end{center}
		\end{figure}
		
		In fact, a very similar pressure-annealing analogy can also be found for undoped CaFe$_2$As$_2$, i.e., in the absence of substitution-induced disorder on the Fe site. To show this, we compare in Fig.\,\ref{fig:Ca122-unified-undoped} the phase diagram of (i) a sample of CaFe$_2$As$_2$ at ambient pressure with $T_{anneal}$ as a tuning parameter \cite{Ran11} and (ii) a sample of CaFe$_2$As$_2$ with $T_{anneal}\,=\,350\,^\circ$C with hydrostatic pressure as a tuning parameter. Note that a sample, which was post-growth annealed at 350$^\circ$C, was found to exhibit the same properties as a sample, grown out of Sn-flux. Thus, despite the difference synthesis route, each sample can be considered to reflect the physical properties of the parent compound \cite{Ran11}. For low hydrostatic pressures or low $T_{anneal}$, CaFe$_2$As$_2$ undergoes a temperature-induced first-order magnetostructural transition at $T_{s,N}$, whereas for high pressures or high $T_{anneal}$ the sample undergoes a transition into a collapsed-tetragonal structure at $T_{cT}$. No indications for superconductivity can be found here, consistent with previous results \cite{Yu09}. For the comparison of $p$ and $T_{anneal}$, the same scaling factor between pressure and annealing temperature ($\Delta T_{anneal}\,=100\,^\circ$C$\,\approx\,\Delta p\,=\,0.0846$\,GPa) was used as for the Co-doped CaFe$_2$As$_2$ samples, the result of which were presented above. By using this conversion factor, a very good matching of the position of the pressure-induced transition from the o/afm to the cT phase can be achieved. In terms of the precise transition temperatures, we find some discrepancies between the ones, inferred from using annealing as a tuning parameter, and the ones from our pressure study close to and in the cT phase. Again, the detailed origin of these discrepancies is presently unknown origin, but a possible explanation might be given by the differences in the strain fields in combination with a phase transition, which involves large lattice changes. Overall, however, on a quite remarkable, quantitative level, postgrowth thermal annealing indeed mimics the effect of hydrostatic pressure for a range of Ca(Fe$_{1-x}$Co$_x$)$_2$As$_2$ samples. Overall, all salient ground states, that are associated with this material class, are readily accessible in this series either in truly hydrostatic pressure conditions or by postgrowth thermal annealing. The data in Figs.\,\ref{fig:Ca122-unified} and \ref{fig:Ca122-unified-undoped} demonstrate that the control of strain in the CaFe$_2$As$_2$ matrix by post-growth-annealing control inclusions has the same effects as the application of hydrostatic pressure for all three of these phases.  Given their different pressure dependencies (size and sign) this strongly suggests that hydrostatic pressure as well as post-growth-annealing control actual strain in very similar manners.

\subsection{Effect of pressure on CaK(Fe$_{1-x}$Ni$_x$)$_4$As$_4$: Occurrence of a new type of half-collapsed tetragonal structure}
	\label{sec:CaK1144}
	
	As opposed to, e.g.,  CaFe$_2$As$_2$, the compound CaKFe$_4$As$_4$ from the family of materials with the 1144 crystal structure is superconducting in its pure, undoped, form, at ambient pressure with a $T_c\,\approx\,35$\,K \cite{Iyo16,Meier16}. The crystal structure of the 1144 family is closely related to the 122 structure, but in contrast to the solid solution Ba$_{1-x}$K$_x$Fe$_2$As$_2$ with I4/$mmm$ body-centered tetragonal symmetry \cite{Wang13}, the 1144 structure has separate, unique crystallographic sites for the Ca and K atoms, resulting in a reduced P4/$mmm$ space group symmetry \cite{Iyo16}. 
	
\begin{figure}[t]%
		\begin{center}
		\includegraphics*[width=0.9\linewidth]{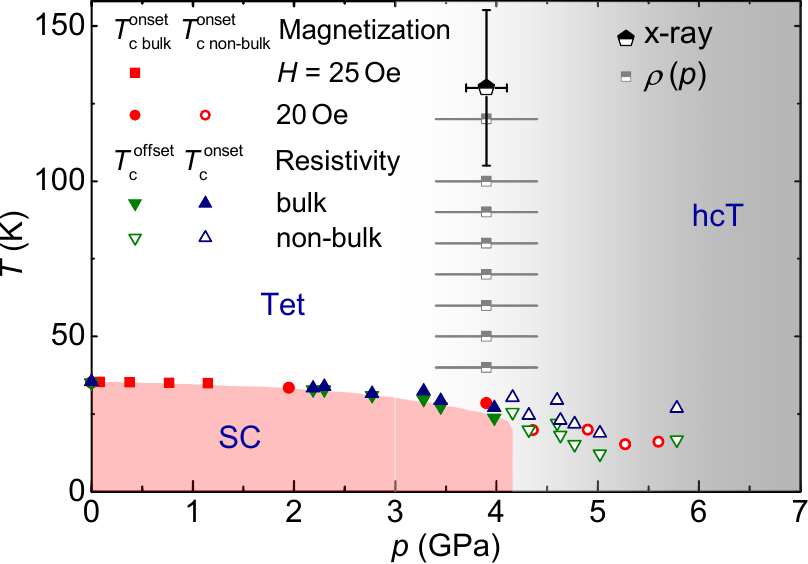}
		\caption{Temperature-pressure phase diagram of CaKFe$_4$As$_4$, determined from measurements of the resistivity (denoted by triangles), the magnetization (squares) and the x-ray diffraction (pentagon) under hydrostatic pressure. The light-red shaded area corresponds to the region of bulk superconductivity (solid symbols), born out of the uncollapsed tetragonal state (white region, denoted as Tet). Grey area indicates the region of half-collapsed tetragonal (hcT) structure in the phase diagram. Reprinted with permission from Ref. \cite{Kaluarachchi17}, Copyright American Physical Society 2017.}
		\label{fig:CaK1144-phasediagram}
		\end{center}
		\end{figure}

	 Followed by the successful growth of single crystals of CaKFe$_4$As$_4$ \cite{Meier16,Meier17}, the effect of pressure on this compound \cite{Kaluarachchi17} was studied by magnetization and resistance measurements in liquid-medium pressure cells and x-ray diffraction measurements under pressure with He-gas as a pressure-transmitting medium up to 6\,GPa. The obtained temperature-pressure phase diagram is shown in Fig.\,\ref{fig:CaK1144-phasediagram}. For low pressures, $p\,\lesssim\,$4\,GPa, robust superconductivity was found with large diamagnetic shielding fraction and zero resistance. In this regime, the superconducting $T_c$ is suppressed, relatively gradually, with increasing pressure. Above 4\,GPa, the superconducting shielding fraction was significantly reduced, whereas the resistance still drops to zero at low temperature. This finding, together with a distinctly different field-dependence below and above 4\,GPa, respectively, indicates a filamentary nature of superconductivity for $p\,\gtrsim\,4$\,GPa, and is very reminiscent of the observations on CaFe$_2$As$_2$, for which a non-bulk superconducting phase was argued to be stabilized by non-hydrostatic pressure components, associated with cooling through the cT transition in a solid medium \cite{Canfield09}. In fact, the x-ray diffraction measurements under pressure identified a pressure-induced structural phase transition close to 4\,GPa, at which the $a$ axis expands by 0.4\,\% and the $c$ axis shrinks by 2.6\,\%. The available data sets suggested that this structural transition is almost vertical at around 4\,GPa in the temperature-pressure phase diagram.
	 
	 The preserved tetragonal symmetry across the pressure-induced structural transition is fully consistent with the idea of a collapsed-tetragonal transition as the origin for the sudden loss of bulk superconductivity. The detailed nature of the structural transition was identified with the help of band structure calculations \cite{Kaluarachchi17,Borisov18}. The results showed that the structural transition at 4\,GPa is a result of the As-As $p_z$ orbital bonding across the Ca-layer, while those orbitals across the K-layer remain unaffected. According to the calculations, the As-As $p_z$ orbitals across the K-layer only bond for significantly higher pressures, $p\,\approx\,12$\,GPa, giving rise to a second structural transition with strong lattice parameter changes (in fact, the predicted second structural transition in CaRbFe$_4$As$_4$ \cite{Borisov18} was subsequently detected \cite{Stillwell19} at high pressures). Correspondingly, the transition at 4\,GPa therefore marks a new type of collapsed tetragonal transition, which was labeled as half-collapsed tetragonal (hcT) transition. It appears very likely that this layer-selective structural collapse in the 1144 family is related to the distinctly different ionic radii of the Ca and the K atom, respectively. This intuition was recently supported by an extensive band-structure calculation study \cite{Borisov18}, which compared the critical pressures for the two hcT transitions for various $AeA$Fe$_4$As$_4$ compounds with different $A$ and $Ae$ species and which established a clear trend of the critical pressures and the cation size. For the impact of this new type of half-collapsed tetragonal transitions on superconductivity, it is important to point out that the first hcT transition already results in a loss of bulk superconductivity in CaKFe$_4$As$_4$. In terms of the magnetism, it was pointed out that for the theoretical description of the pressure-induced hcT transitions, it is important to simulate the presence of spin-vortex fluctuations by imposing a ''frozen'' spin-vortex configuration on the Fe sites \cite{Borisov18}. In these calculations, the ''magnetic collapse'', which is naturally absent in real crystals of CaKFe$_4$As$_4$, occurs in close proximity to the first hcT transition. Only for the special cases of Eu-containing 1144 compounds, such as CsEuFe$_4$As$_4$ and RbEuFe$_4$As$_4$, the ferromagnetism, which is associated with the Eu$^{2+}$ moments, survives the hcT transitions, as consistently found in calculations \cite{Borisov18} and experiments \cite{Jackson18,Xiang19}.
	 
	 		\begin{figure}[t]%
		\begin{center}
		\includegraphics*[width=0.75\linewidth]{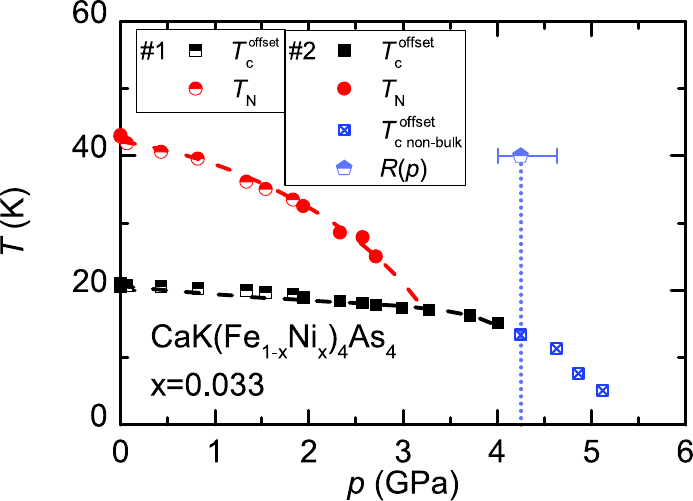}
		\caption{Temperature-pressure phase diagram of CaK(Fe$_{1-x}$Ni$_x$)$_4$As$_4$ with $x\,=\,0.033$, constructed from high-pressure resistance measurements. Red circles (black squares) correspond to the antiferromagnetic transition at $T_N$ (the superconducting transition at $T_c$). Blue squares indicate the transition into a filamentary superconducting state. Blue pentagon indicates the position of an anomaly, which is likely related to a pressure-induced structural transition from the uncollapsed tetragonal to the half-collapsed tetragonal structure. Reprinted with permission from Ref. \cite{Xiang18}, Copyright American Physical Society 2018.}
		\label{fig:Nidoped-CaK1144-phasediagram}
		\end{center}
		\end{figure}

	The robustness of these ideas concerning the relation of the half-collapsed tetragonal transition and superconductivity can be seen, e.g., when considering the temperature-pressure phase diagram of electron-doped CaKFe$_4$As$_4$. At ambient pressure, the magnetic ground state, which was observed in the series of CaK(Fe$_{1-x}$Ni$_x$)$_4$As$_4$ and CaK(Fe$_{1-x}$Co$_x$)$_4$As$_4$, has created enormous interest, since it was shown that the magnetic order is of a new type, the so-called hedgehog spin-vortex order \cite{Meier18}, which is likely stabilized by the reduced symmetry of the 1144 crystal structure. Given the existence of two inequivalent As sites in this structure (see Fig.\,\ref{fig:overviewfigure}\,(a)), this magnetic order features an alternating all-in and all-out motif around the As(1) sites, whereas the As(2) sites do not manifest any transferred magnetic hyperfine field, and thus preserves the high-temperature tetragonal symmetry \cite{Meier18,Kreyssig18}. For the discussion of the effect of hydrostatic pressure on this doped CaKFe$_4$As$_4$ system, we show in Fig.\,\ref{fig:Nidoped-CaK1144-phasediagram} exemplarily the temperature-pressure phase diagram of a single crystal of CaK(Fe$_{1-x}$Ni$_x$)$_4$As$_4$ with $x\,=\,0.033$, constructed from high-pressure resistance measurements up to 6\,GPa in a pressure cell with a liquid as pressure-transmitting medium \cite{Xiang18}. The studied sample does not only depict the new type of hedgehog spin-vortex order, but also shows superconducting order with $T_c\,<\,T_N$. The transition temperatures of both orders are suppressed with increasing pressure, until for $p\,\geq\,3\,$GPa only the superconducting transition could be detected. Increasing pressure beyond 4\,GPa results in a sudden change of the ground state: only indications of a filamentary superconducting state could be found. At the same time, the pressure dependence of the resistance strongly suggested that the single crystal undergoes a pressure-induced structural phase transition close to 4\,GPa. Based on the analogy to undoped CaKFe$_4$As$_4$ \cite{Kaluarachchi17}, it seems thus very likely that CaK(Fe$_{1-x}$Ni$_x$)$_4$As$_4$ with $x\,=\,0.033$ is tuned through a hcT transition at 4\,GPa. The fact that the critical pressure of the hcT transition is almost unchanged with Ni substitution is not at all surprising, given the fact that the hcT transition is associated with the bonding of As orbitals across the Ca-layer \cite{Borisov18}.
		
		In terms of the interplay of the hedgehog spin-vortex magnetism and superconductivity, it is remarkable to point out that both the magnetic transition at $T_N$ and the superconducting transition at $T_c$ are both suppressed with increasing $p$ in CaK(Fe$_{1-x}$Ni$_x$)$_4$As$_4$ with $x\,=\,0.033$. This is in contrast to the observations on many other iron-based superconductors, for which either doping or pressure results in increase of $T_c$, when $T_N$ is suppressed. This opposite trend of the transition temperatures is commonly considered as a signature of competing orders. In order to reconcile the observations on CaK(Fe$_{1-x}$Ni$_x$)$_4$As$_4$ with the scenario of competing orders, we would like to return to an argument, which we brought up in Sec.\,\ref{sec:FeSe} in the discussion of FeSe on the basis of an itinerant model for competing superconducting and spin-density wave order \cite{Machida81}. In the picture of competing orders, both orders are allowed to be suppressed with pressure, as long as $|$d$T_M/$d$p|\,>\,|$d$T_c/$d$p|$, when superconductivity is the order, which is promoted by the application of pressure. This then gives rise to an ''effective'' suppression of $T_M$ with respect to $T_c$. This condition is indeed satisfied for pressure-tuned Ni-doped CaKFe$_4$As$_4$. The competition of spin-vortex magnetism and superconductivity was recently also confirmed by microscopic M\"ossbauer measurements \cite{Budko18} on a sample with this particular Ni concentration at ambient pressure, as well as over wider $x$ ranges of the series of CaK(Fe$_{1-x}$Ni$_x$)$_4$As$_4$ samples by M\"ossbauer \cite{Budko18} and neutron \cite{Kreyssig18} measurements at ambient pressure. That magnetism and superconductivity clearly interact, can also be seen from the fact that the slope $|$d$T_c$/d$p|$ becomes distinctly larger around 3\,GPa, i.e., right when the magnetism is absent. 

\subsection{Superelasticity}
\label{sec:superelasticity}

The structural collapse, which we discussed in the two previous section for CaFe$_2$As$_2$ and CaKFe$_4$As$_4$ in terms of its consequence on electronic properties, was recently also found to be responsible for extraordinary elastic properties under uniaxial compression. In contrast to most intermetallic compounds, which, owing to their brittleness, usually allow for a maximum of pressure-induced strain of less than 1\,\% before the fracture, micropillars (with dimensions of 2\,$\mu$m in diameter and 6\,$\mu$m in height) of the above-mentioned pncitides were found to exhibit up to 17\,\% of recoverable strain, by tuning via uniaxial stress through the collapsed tetragonal (cT) and the half-collapsed tetragonal (hcT) transitions, respectively \cite{Sypek17,Song19}. Given the strong interrelation of the electronic properties and this structural distortion, these materials thus form a very promising platform for strain engineering, in which, e.g., superconductivity can be switched on and off through the superelasticity process \cite{Song19}.

\section{Role of uniaxial strain for probing and tuning electronic order in iron-based superconductors}
\label{sec:uniaxialstrain}

		\begin{figure*}[t]%
		\begin{center}
		\includegraphics*[width=0.75\textwidth]{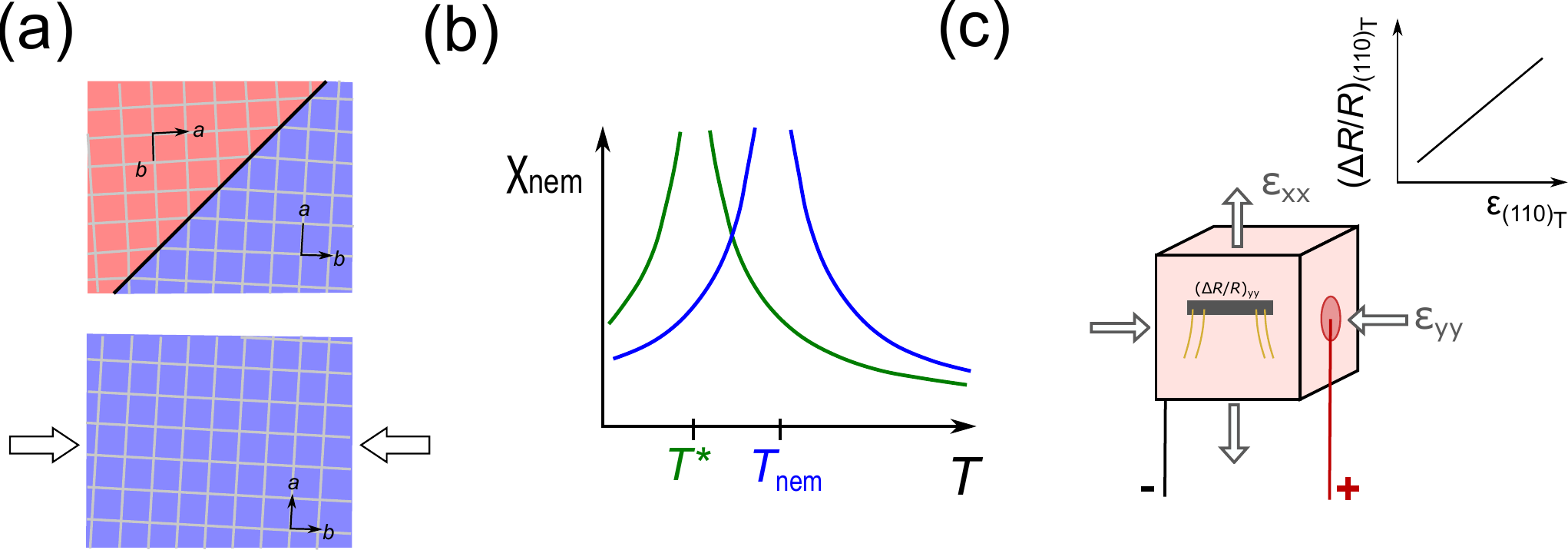}
		\caption{Role of uniaxial pressure and strain for probing the properties of the nematic state. (a) Detwinning the orthorhombic state. Upon cooling through the tetragonal-to-orthorhombic transition, structural twin boundaries are formed over the full crystal (top). When a sufficiently large uniaxial pressure is applied to the system prior to cooling through the structural transition, a single-domain state is realized (bottom); (b) and (c) Determination of the nematic susceptibility. (b) In a Ginzburg-Landau mean-field model, the temperature dependence of the nematic susceptibility $\chi_{nem}$ follows a Curie-Weiss temperature dependence. If there is no coupling of the electronic, nematic order parameter to the lattice, the electronic system undergoes the nematic transition at $T^\star$. Any finite, bilinear coupling to the lattice renormalizes the nematic susceptibility, and the corresponding transition temperature is increased to a temperature $T_{nem}\,>\,T^\star$ (after \cite{Boehmer16d}); (c) Experimentally, the nematic susceptibility for $T\,>\,T_{nem}$ can be determined by measurements of the resistance change, induced by changes of the strain along the symmetry-breaking crystallographic direction. For BaFe$_2$As$_2$, this direction corresponds to the (110)$_T$ direction. The slope of the linear relation of ($\Delta R/R$) vs. $\epsilon$ (inset) is directly proportional to $\chi_{nem}$. To obtain this relation, a sample of interest is typically fixed to a piezoelectric actuator, which can be strained \textit{in situ} by the application of a voltage.}
		\label{fig:role-uniaxial}
		\end{center}
		\end{figure*}

Uniaxial mechanical stress has become an essential tool to (i) probe the properties of the nematic phase in iron-based superconductors, but also (ii) to tune the transition temperature of all salient ground states that are associated with iron-based superconductors. In the following, we will shortly summarize the use of uniaxial stress in both regards.

For a study of the in-plane anisotropy of the nematic-orthorhombic phase of iron-based superconductors, uniaxial stress is essential, since the tetragonal-to-orthorhombic transition results in the formation of structural twin boundaries, i.e., an alternation of $a$ and $b$ axis, which leads to an averaged, quasi-isotropic response for macroscopically measured quantities (see Fig.\,\ref{fig:role-uniaxial}\,(a)). Uniaxial stress is, thus, used for detwinning the orthorhombic state, by imposing a preferential orientation for the formation of $a$ and $b$ axis, respectively \cite{Chu10,Tanatar10,Boehmer15b}. In this way, the anisotropic in-plane response below the structural transition can be revealed. Measurements of the in-plane resistivity anisotropy \cite{Chu10,Tanatar10}, performed by using this method of detwinning, on Ba(Fe$_{1-x}$Co$_x$)$_2$As$_2$ were early on interpreted as strongly indicative of an electronic origin of the structural distortion. Specifically, the so-revealed electronic anisotropy $\rho_b/\rho_a$, with $\rho_a$ ($\rho_b$) the resistivity along the longer $a$ (shorter $b$) axis, was found to depend non-monotonically on doping and disorder, and in particular does not correlate with the evolution of the structural anisotropy, $\frac{a-b}{a+b}$ \cite{Chu10}. This view of an electronic origin of nematicity was supported early on by theoretical considerations \cite{Fernandes14}. Similarly, many other measurement quantities were found to display a directional anisotropy in the orthorhombic state in various materials \cite{Yi11,Fu12,Kasahara12,Jiang13,Nakajima11,Mirri15,He17,Kissikov18,Baum18,Pfau19}. In particular, an in-plane anisotropy \cite{Fu12} in the magnetic susceptibilities as well as in the on-site energies of the $d_{xz}$ and $d_{yz}$ orbitals \cite{Yi11} were found experimentally to set in at $T_s$, manifesting the intimate interplay of structural, magnetic and orbital order in the nematic state.

More recently, efforts were extended to study the characteristics of the nematic fluctuations in the tetragonal state, out of which the low-temperature ordered state is born. This allows for the study of the tendency of a system towards nematic order even in the absence of long-range order, e.g., close to potential quantum-critical points. In general, any order-parameter susceptibility can be accessed by measuring the response of a system to small changes of the conjugate field. Since nematicity is coupled to an orthorhombic distortion, uniaxial strain along the symmetry-broken crystallographic direction serves as a conjugate field in the case of nematic transitions \cite{Chu12,Fernandes14}. Correspondingly, measurements under small uniaxial strain in the tetragonal state allow for the inference of the nematic susceptibility. For this purpose, a proxy for the unknown nematic order parameter has to be used. Commonly, the electronic resistivity anisotropy is assumed to be a good proxy for the nematic order parameter. Based on these assumptions, it then follows that the change of the resistivity anistropy with uniaxial strain, which can be determined in elastoresistance measurements using the novel piezo-based devices (see Fig.\,\ref{fig:role-uniaxial}\,(c)), introduced in Sec.\,\ref{sec:experimentalmethods}, is proportional to the nematic susceptibility \cite{Chu12,Kuo13,Kuo16}. Experimentally, this technique was initially employed for the series of BaFe$_2$As$_2$ and Ba(Fe$_{1-x}$Co$_x$)$_2$As$_2$ \cite{Chu12,Kuo16}, with strain applied along the tetragonal (110)$_T$ direction, which either becomes the orthorhombic (100)$_O$ axis or the (010)$_O$ axis, respectively, upon cooling through $T_s$. One of the main results was that, whenever the system undergoes a structural transition at a finite temperature $T_s$, the nematic susceptibility almost diverges when approaching the structural transition from above \cite{Chu12,Kuo13,Kuo16}. This result, again, was considered as a very strong evidence that the orthorhombic distortion is not the primary order parameter itself, but rather a secondary cause of an electronic order parameter with unknown microscopic origin. If the orthorhombic distortion was to be the primary order parameter, the resistivity anisotropy would simply be proportional to the orthorhombic distortion \cite{Fernandes14}, and therefore not display a strong temperature dependence above $T_s$. In more detail, it was found that the temperature dependence of the nematic susceptibility follows a Curie-Weiss law (see Fig.\,\ref{fig:role-uniaxial}\,(b)). By employing a Ginzburg-Landau ansatz, which includes a symmetry-allowed bilinear coupling term $\lambda$ between the electronic order parameter and the lattice strain, this behavior was rationalized and the Curie constant $C$ assigned to the strength of the bilinear coupling, and the Curie-Weiss\ temperature $T^\star$ to the bare electronic nematic transition temperature (i.e., in the absence of a coupling to the crystal lattice). The underlying notion is that the nematic fluctuations make the crystalline lattice soft, which results for $\lambda\,>\,0$ in a structural transition at $T_s\,>\,T^\star$ and simultaneously a nematic transition, as expressed in a peak of the nematic susceptibility at $T_s$ \cite{Boehmer16d}.

Followed by this important finding, the nematic susceptibility was studied by elastoresistance measurements in different iron-based superconductors across wide ranges of their phase diagrams. Measurements on P-, Ni-, Co- and K-substituted BaFe$_2$As$_2$ as well as Te-substituted FeSe all revealed a diverging nematic susceptibility \cite{Kuo16} near their respective critical dopings as as a function of temperature. Based on this result, it was suggested that the divergence of the nematic susceptibility might be a generic feature or iron-based superconductors and might indicate the presence of an underlying nematic quantum-critical point. Further, it was suggested that superconducting pairing might be influenced by the underlying nematic quantum-critical point, and thus the potential promotion of superconductivity by nematic fluctuations deserves further considerations. Similar conclusions were also inferred for the series of Fe(Se$_{1-x}$S$_x$) \cite{Hosoi16}, for which the nematic transition can be tuned to zero in the absence of magnetism.

The above-discussed cases of nematic order have Ising symmetry, and thus, the nematic order is locked by the crystallographic directions. Very recently, elastoresistance measurements, which can also be used to infer symmetry information on the underlying nematic order parameter, also suggested that a $XY$-type of nematic order, i.e., a nematic order, which can point in any arbitrary direction, might be realized in a narrow doping window in Ba$_{1-x}$Rb$_x$Fe$_2$As$_2$ \cite{Ishida19}. The end members of this series show a nematic ground state with so-called $B_{2g}$ symmetry and one with $B_{1g}$ symmetry, respectively, and thus the new $XY$-type nematic order was proposed to form in between these two extreme limits. From band structure calculations \cite{Borisov19b}, it was argued that the change of the nematic order with so-called $B_{2g}$ symmetry in BaFe$_2$As$_2$ to $B_{1g}$ symmetry in RbFe$_2$As$_2$ is related to the change of the magnetic ground state configuration from single stripe to double stripe.

The studies of elastoresistance were also very recently extended to explore nematic degrees of freedom in other material classes. For example, in a work on cuprate superconductors, enhanced nematic fluctuations were reported close to the pseudogap critical endpoint \cite{Orth19} based on elastoresistance measurements \cite{Ishida19b}. Overall, driven by efforts on the iron-based superconductors \cite{Chu12,Kuo13,Watson15,Tanatar16,Kuo16,Palmstrom17,Hristov18,Hong19,Palmstrom19,Hristov19,Straquadine19}, elastoresistance measurements \cite{Shapiro15,Walmsley17} have emerged as an important tool for the study of nematicity  within a short time, and have only begun to be applied to the wider class of correlation electron systems, for which the relevance of nematic degrees of freedom is nowadays appreciated \cite{Keimer15,Fradkin10,Fernandes14,Fernandes19,Seo20}. In addition, the research on the iron-based superconductors has driven initiatives to gain further insights into the intriguing properties of the nematic state by exploring a variety of strain-derivatives of physical properties, such as the Seebeck and Nernst effect \cite{Caglieris19}, and will likely motivate further theoretical works of understanding the overall strain response of transport quantities \cite{Jo19}.

Beyond probing the properties of the nematic state and its fluctuations through a control of uniaxial strain and stress, uniaxial stress can also be used to tune phase transitions. We have already introduced a particularly clear example of uniaxial stress tuning of iron-based superconductors above in Sec.\,\ref{sec:superelasticity}, where we discussed how uniaxial strain has led to a remarkably large recoverable strain, associated with the collapsed-tetragonal transition in CaFe$_2$As$_2$ and CaKFe$_4$As$_4$ and the concomitant loss of bulk superconductivity.

The tunability of the nematic and magnetic phase transition by uniaxial pressure has been tested in the example case of an underdoped Ba(Fe$_{1-x}$Co$_x$)$_2$As$_2$ \cite{Ikeda19}. This response was carefully compared to the hydrostatic pressure response. In this way, it was possible to perform a symmetry decomposition of the strains, which are induced by the uniaxial stress. As a result, the separate response of the nematic transition to symmetric and antisymmetric strains, which are both unavoidably present when a sample is exerted to uniaxial pressure, was inferred. It turned out that in particular antisymmetric strain of $B_{2g}$ symmetry might be a particularly suitable parameter to tune materials across a nematic critical point. In a very recent posting \cite{Worasaran20}, this deeper understanding of the tuning of the nematic transition by strains of different symmetries was used to discuss the quantum-critical nature of the nematic fluctuations. In addition, the response of the superconducting state in Ba(Fe$_{1-x}$Co$_x$)$_2$As$_2$ to anisotropic strain was studied experimentally very recently \cite{Malinowski19}. The main finding of Ref. \cite{Malinowski19} is that both compressive as well as tensile strain suppress the superconducting $T_c$ very quickly, and can even induce a superconductor-to-metal quantum phase transition. These observations were assigned to the competition of superconductivity and the magnetic-nematic state, since the latter is promoted by the application of anisotropic strain.

For completeness, we would like to mention at this point that strain can not only be applied uniaxially, but in principle also biaxially. Here, the sample is strained in two crystallographic direction simultaneously. The amount and direction of strain (compressive or tensile) can be either the same or different along the two different directions. Experimentally, such kind of strain is usually achieved nowadays from a rigid gluing of samples to a substrate by utilizing the difference in temperature-dependent thermal expansion, see e.g. \cite{Boehmer17,He17,Ran14}. For systems with low crystallographic symmetry, this procedure unavoidably results in an anisotropic biaxial strain, even if the substrate shows an isotropic expansion, since the thermal expansion of the sample is anisotropic. For the iron-based superconductors, studied here, the tetragonal symmetry of the crystal structure ensures that isotropic biaxial strain can be achieved by a rigid gluing of samples and substrate, when the sample is strained along the two in-plane directions and the substrate shows an isotropic expansion. This type of biaxial strain directly tunes the $c/a$ ratio of tetragonal samples, and thus is in its effect very similar to uniaxial pressure along the crystallographic $c$ direction, which is often difficult to achieve experimentally because of the small out-of-plane dimensions of crystals of many tetragonal intermetallic systems. This different tuning ansatz of biaxial strain was recently established for the iron-based superconductor series Ca(Fe$_{1-x}$Co$_x$)$_2$As$_2$ \cite{Boehmer17,Ran14}, the hydrostatic pressure response of which was discussed here in Sec.\,\ref{sec:Ca122}. Given the high sensitivity of this series to changes of pressure and strain, it turned out that biaxial in-plane strain, and thus the $c/a$ ratio, is a very efficient tuning parameter for the phase diagram of Ca(Fe$_{1-x}$Co$_x$)$_2$As$_2$. It is important to note, that in contrast to, e.g., hydrostatic pressure experiments, here the strain and not the applied pressure is the control parameter. In particular when a first-order structural transition is involved, such as the o/afm transition in Ca(Fe$_{1-x}$Co$_x$)$_2$As$_2$, this might lead to a partial strain release upon cooling through the phase transition and therefore results in a well-defined phase coexistence. As a consequence, attention has to be paid when fixing thin samples rigidly to a substrate, as commonly employed in various measurements. It becomes particularly important, when the system is highly sensitive to pressure and strain: whereas the high pressure sensitivity does not only offer the great possibility to tune materials conveniently in laboratory experiments, it can, if care is not taken, also result in unwanted modifications related to the exact mounting of the sample for experiments \cite{Canfield16,Boehmer17,Ran14}.   

\section{Combining hydrostatic and uniaxial pressure: Proof-of-principle and future perspective}
\label{sec:hydrostaticanduniaxial}

		\begin{figure}[t]%
		\begin{center}
		\includegraphics*[width=0.7\linewidth]{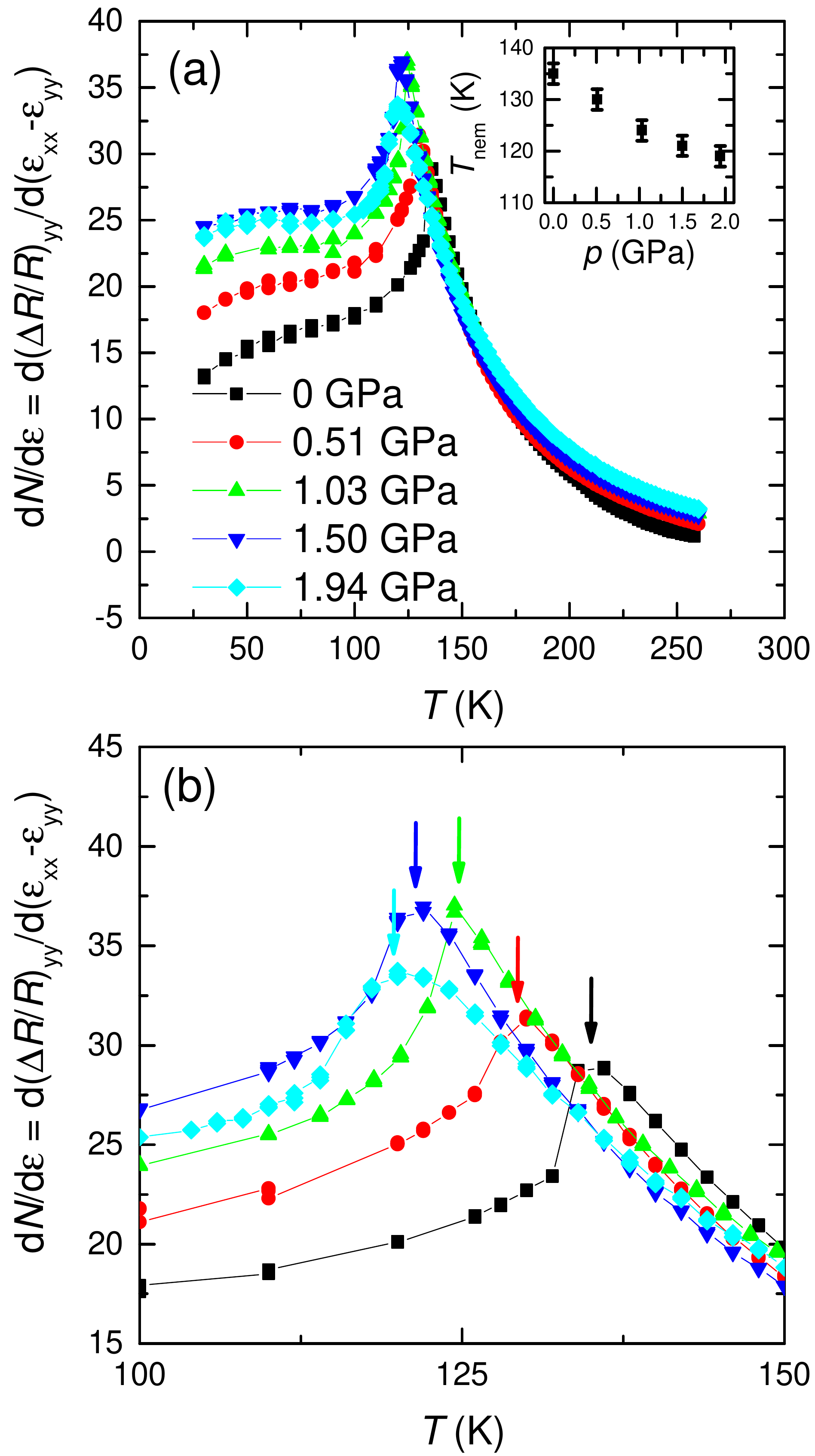}
		\caption{Elastoresistance, d$(\Delta R_{yy}/R_{yy})$/d$(\epsilon_{xx}-\epsilon_{yy})$, under hydrostatic pressure (0\,GPa$\,\le\,p\,\le\,$1.94\,GPa) on BaFe$_2$As$_2$ over the full temperature range 30\,K$\,\le\,T\,\le\,$300\,K (a) and an enlarged view of the elastoresistance close to the peak position at $T_{nem}$ (b). Inset of (a) shows the evolution of the $T_{nem}$ with hydrostatic $p$, inferred from the peak position in d$(\Delta R_{yy}/R_{yy})$/d$(\epsilon_{xx}-\epsilon_{yy})$ (see arrows). Reprinted with permission from Ref. \cite{Gati20}, Copyright AIP Publishing 2020.}
		\label{fig:Ba122-elasto-pressure}
		\end{center}
		\end{figure}	

The importance of hydrostatic and uniaxial pressure for probing and tuning the ground states of the family of iron-based superconductors makes it very compelling to explore the option of combined hydrostatic and uniaxial pressure experimentally, and to use the iron-based superconductors as test cases to explore the impact of this novel tuning combination. To this end, we developed a miniaturized version of the piezo-based strain device, which was initially introduced by the Stanford group for measurements of the elastoresistivity \cite{Chu12,Kuo13,Kuo16}. Our modified version is small enough to fit into a conventional piston-pressure cell. For the details of the experimental setup we refer the reader to Ref. \cite{Gati20}. Instead, we first want to stress that our main finding of Ref. \cite{Gati20} is that the piezoelectric actuators, which are used to strain samples \textit{in situ} in a quasi-uniaxial manner (given their highly anistropic biaxial expansion, see below for more details), can demonstrably operate over wide ranges of pressure (up to $\approx\,2$\,GPa) and temperature (low $T$ up to room temperature). This result is important since it cannot be taken for granted, as (i) the piezoelectric actuator might break, if exposed to large pressures, (ii) it might not be able to act against the significant external forces that are excerted on it by the medium, in particular when the pressure medium becomes solid \cite{Torikachvili15} or (iii) it might be not possible to apply large-enough voltages to drive the actuator via the pressure-cell feedthrough. In fact, it turned out that none of this three potential issues are significant enough to prevent usage: (i) no breakage or visual damage was found in the actuators in an inspection after a pressure cycle; (ii) a clear and measurable strain, induced by the external voltage, was determined from \textit{in situ} measurements of a strain-gauge resistance for all temperatures and pressures investigated, even when the medium was solid and (ii) no voltage ''breakdown'' was observed up to 150\,V, the highest voltage used in these experiments. 

				\begin{figure*}[t]%
		\begin{center}
		\includegraphics*[width=0.9\textwidth]{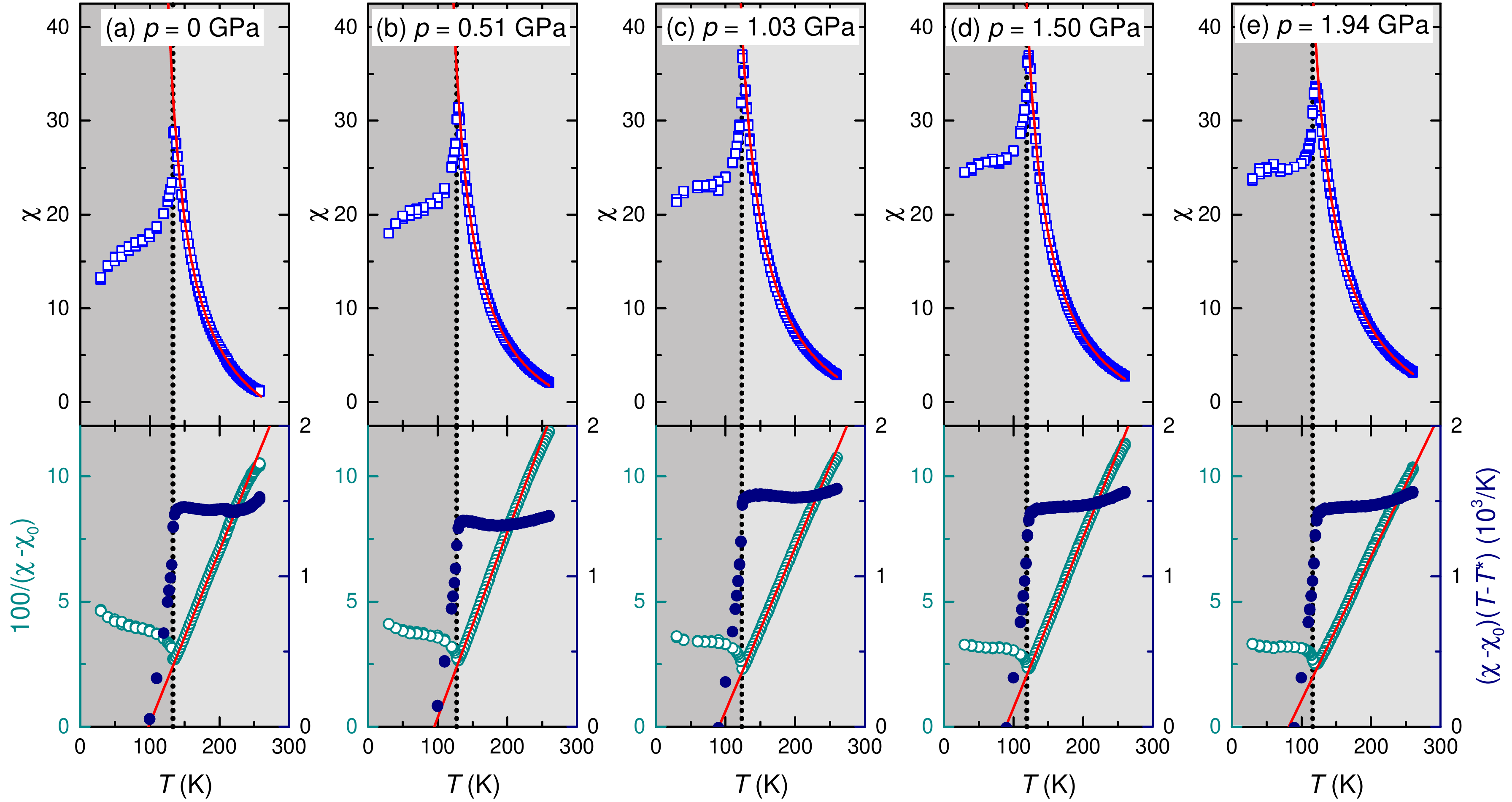}
		\caption{Temperature ($T$)-dependent nematic susceptibility of BaFe$_2$As$_2$ for different pressures in the range 0\,GPa$\,\le\,p\,\le\,$1.94\,GPa ((a)-(e)), as extracted from measurements of the elastoresistance; Top panel shows the measured elastoresistance $\chi\,=\,$d$(\Delta R_{yy}/R_{yy})$/d$(\epsilon_{xx}-\epsilon_{yy})$ (symbols), which is related to the nematic susceptibility, $\chi_{nem}$ via $\chi\,\simeq\,\chi_{nem} + \chi_0$, with $\chi_0$ being a parameter, which describes effects not related to nematicity. $\chi_{nem}$ is expected to follow a Curie-Weiss-type temperature dependence, $\chi_{nem}\,=\,\frac{C}{T-T^\star}$ (see text for details). The red line corresponds to such a Curie-Weiss fit of the experimental data. The bottom panel depicts the same data as in the top panel in different representations. A plot of the inverse nematic susceptibility $\frac{1}{\chi-\chi_0}$ (open symbols) is shown on the left axis, a plot of the Curie constant $C$, as calculated by $(\chi-\chi_0)(T-T^\star)$, on the right axis. Reprinted with permission from Ref. \cite{Gati20}, Copyright AIP Publishing 2020.}
		\label{fig:Ba122-Curiefitting}
		\end{center}
		\end{figure*}

For our proof-of-principle demonstration that our setup allows for the exploration of the combined effects of hydrostatic and quasi-uniaxial pressure via elastoresistance measurements under pressure, we chose BaFe$_2$As$_2$ as an example case, given that (i) its temperature-dependent elastoresistance behavior is well-characterized at ambient pressure \cite{Chu12,Kuo13,Kuo16} and (ii) the details of the temperature-pressure phase diagram up to 2\,GPa have been explored in detail by a variety of techniques (see Sec.\,\ref{sec:Ba122} \cite{Gati19c}). For this purpose, a sample, which was cut along the (110)$_T$ direction was mounted on the piezoelectric actuator. When this actuator is strained by the application of a voltage, the actuator is, strictly speaking, strained in a biaxial manner, though the biaxial strain is very anisotropic. Whereas the actuator expands in a direction, which we denote by $x$, giving rise to an additional strain $\epsilon_{xx}$ on the sample, it compresses in the other direction, which results in a negative strain $\epsilon_{yy}$. Thus, the effective strain, which the sample experiences along the (110)$_T$, is quantified by $\epsilon_{xx}-\epsilon_{yy}$. The studied sample was mounted in the $yy$-direction and its resistance is therefore referred to $R_{yy}$. The measured elastoresistance, d$(\Delta R_{yy}/R_{yy})$/d$(\epsilon_{xx}-\epsilon_{yy})$, serves as a proxy for the change of the resistance anisotropy with strain (details on the underlying symmetry considerations etc. can be found in detail in Refs. \cite{Kuo13,Kuo16,Gati20}), and is shown in Fig.\,\ref{fig:Ba122-elasto-pressure}\,(a) for different pressures in the range 0\,GPa$\,\le\,p\,\le\,$1.94\,GPa. The lowest pressure data, which is labelled with 0\,GPa, was taken in inside the pressure cell without the application of an external force, giving rise to zero pressure at low temperatures. A comparison of this specific data set with literature results, taken at ambient pressure, \cite{Chu12,Kuo13,Kuo16} has shown an excellent agreement for high temperatures in the tetragonal state. Upon lowering the temperature from high temperatures, d$(\Delta R_{yy}/R_{yy})$/d$(\epsilon_{xx}-\epsilon_{yy})$ displays a strong increase, until it reaches a maximum, which we denote as the nematic transition temperature $T_{nem}(p\,=\,0)\,\approx\,135\,$K and which coincides with the structural transition temperature $T_s(p\,=\,0)$ \cite{Canfield10}. Below $T_{nem}$, d$(\Delta R_{yy}/R_{yy})$/d$(\epsilon_{xx}-\epsilon_{yy})$ drops quickly and flattens upon further cooling. In this temperature range, the elastoresistance response is dominated \cite{Hristov19b} by the formation of the structural twin domains, and thus is expected to be dominated by extrinsic effects. Upon pressurization, the overall temperature dependence of d$(\Delta R_{yy}/R_{yy})$/d$(\epsilon_{xx}-\epsilon_{yy})$ is not strongly affected by pressure. For all pressures, the elastoresistance increases strongly upon cooling down to a pressure-dependent peak position $T_{nem}(p)$ and quickly drops below this temperature, until the elastoresistance is basically temperature-independent soon after cooling through $T_{nem}(p)$. A closer look on the shift of the peak position with pressure is shown in Fig.\,\ref{fig:Ba122-elasto-pressure}\,(b). The peak position, which is associated with the nematic transition temperature $T_{nem}$, is shifted to lower temperatures with increasing pressure with a rate of $\approx\,-(8.5\,\pm\,1)\,$K/GPa. A decrease of the nematic and structural transition temperatures with pressure is expected based on previous studies, and in fact, the suppression rate, revealed here, is even on a quantitative level consistent with the ones of free-standing samples \cite{Gati20}.
		
		For a quantitative evaluation of the evolution of the nematic susceptibility $\chi_{nem}$ with pressure, we performed a fitting of the experimental data set, shown in Fig.\,\ref{fig:Ba122-elasto-pressure}, by a modified Curie-Weiss law \cite{Chu12,Kuo13,Kuo16} and present it in Fig.\,\ref{fig:Ba122-Curiefitting}. As mentioned above, this approach is motivated by a Ginzburg-Landau ansatz, which takes into account a bilinear coupling between the electronic, nematic order parameter and the lattice strain. The modified Curie-Weiss law has the form $\chi_{nem}+\chi_0\,=\,\frac{C}{T-T^\star}+\chi_0$, with $\chi_0$ a free parameter, which takes temperature-independent contributions to the elastoresistance into account, that are not related to nematicity, and with $C$ and $T^\star$ being the Curie constant and the bare electronic transition temperature, as defined in the previous Sec.\,\ref{sec:uniaxialstrain}, respectively. The fitting of our experimental data for all five measured pressures is shown in the top panels of Fig.\,\ref{fig:Ba122-Curiefitting}\,(a). The very good description of our experimental data with this modified Curie-Weiss law can not only be seen from the raw data in the upper panels of (a), but also from a plot of the inverse nematic susceptibility $1/\chi_{nem}\,=\,1/(\chi-\chi_0)$, with $\chi$ being the measured elastoresistance d$(\Delta R_{yy}/R_{yy})$/d$(\epsilon_{xx}-\epsilon_{yy})$, shown in the lower panels of Fig.\,\ref{fig:Ba122-Curiefitting}\,(b). The latter plots also visualize that the bare electronic transition temperature $T^\star$ is suppressed with increasing pressure. In addition, a plot of $(\chi-\chi_0)(T-T^\star)$, also shown in the lower panels in Fig.\,\ref{fig:Ba122-Curiefitting}\,(b), shows how the Curie constant $C$ remains, within the error bars, almost unchanged with increasing pressure. Overall, this analysis leads to the conclusion that nematic fluctuations prevail in BaFe$_2$As$_2$ under pressures up to at least 2\,GPa without a drastic change of the electron-lattice coupling strength. More importantly, these results serve as a proof-of-principle example that elastoresistance can be measured under pressure to explore the evolution of the nematic susceptibility using a tuning parameter, which is very fine-tunable and not subject to changing levels of disorder. As such, it will be possible, in the future, to study the nematic susceptibility across pressure-tuned nematic quantum-critical points, such as in Co-doped BaFe$_2$As$_2$ \cite{Canfield10}. Even further, the setup, presented here, enables to study the combined action of hydrostatic and quasi-uniaxial pressure, and thus, might be of relevance for tuning and probing the properties of the wider class of correlated electron materials, whenever they are coupled to the crystalline lattice.

\section{Summary and Perspectives}
\label{sec:summary}

In this article, we have presented recent experimental progress in tuning and probing the various ground states of iron-based superconductors by pressure. Given that members of this material class are known for their complex interplay of electronic and structural degrees of freedom, they represent a particularly suitable playground to develop a systematic understanding of the impact of different types of pressure tuning, i.e., uniaxial, biaxial and hydrostatic pressure, on the electronic properties. Since all these tuning parameters do not involve inherently changing the level of disorder, compared to e.g. substitution studies, experimental studies under pressure offer the great opportunity for a better comparison with theoretical models. In this article, we focused on elucidating the fascinating interplay between superconductivity, magnetism, nematicity and structural collapsed transitions under pressure on selected iron-based superconductor systems. In particular, we discussed (i) an unusual behavior of the superconducting and magnetic transition in pressurized FeSe, giving rise to wide temperature ranges of non-long range and non-static electronic orders, (ii) the impact of a pressure-induced change of the Fermi surface topology in underdoped Ba(Fe$_{1-x}$Co$_x$)$_2$As$_2$ on the evolution of nematic order with respect to magnetic order, (iii) the increase of the superconducting critical $T_c$ in Ba(Fe$_{1-x}$Co$_x$)$_2$As$_2$ close and beyond optimal doping by hydrostatic pressure, (iv) the impact of the different types of structural collapsed transition in Ca(Fe$_{1-x}$Co$_x$)$_2$As$_2$ and CaK(Fe$_{1-x}$Ni$_x$)$_4$As$_4$ on the electronic properties, and (v) a study of the nematic susceptibility under hydrostatic pressure by straining BaFe$_2$As$_2$ \textit{in situ} inside the pressure cell. The presented efforts were, to a large extent, driven by the development of new techniques to be readily available under pressure. This includes, for example, an $ac$ calorimetric technique, which allows for the study of the specific heat - a crucial tool for the determination of phase diagrams - with large sensitivity over wide temperature and pressure ranges. (We have shown recently by studying the temperature-pressure phase diagram of a magnetic van-der-Waals material \cite{Gati19d} that this technique is and will be of use for the wider class of correlated electron systems, whenever they are susceptible to pressure tuning.) In addition, we also presented our recent efforts in combining hydrostatic and uniaxial pressure. This novel ansatz has a direct relevance for the study of nematicity across pressure-tuned nematic critical points, but might also offer a novel way of tuning correlated electron materials in general.

\section{Acknowledgments}
We thank Michael Lang, Stephan Kn\"oner, Sebastian K\"ohler and Bernd Wolf for the fruitful collaboration on the measurements, performed under He-gas pressure. We also thank Roser Valent\'{i}, Anna E. B\"ohmer, Vladislav Borisov, Udhara Kaluarachchi, Andreas Kreyssig, Valentin Taufour, Milton Torikachvili and Rafael Fernandes for fruitful collaborations and discussions over the years. EG thanks Gil Drachuck for his assistance in setting up the specific heat setup, discussed in this paper. Finally, we thank Michael Lang (EG), Aleksander G. Gapotchenko and Efim S. Itskevich (SLB) and Joe Thompson (PCC) for having introduced us to the use of pressure as a measurement parameter. Work at the Ames Laboratory was supported by the U.S. Department of Energy, Office of Science, Basic Energy Sciences, Materials Sciences and Engineering Division. The Ames Laboratory is operated for the U.S. Department of Energy by Iowa State University under Contract No. DEAC02-07CH11358. E.G. and L.X. were funded, in part, by the Gordon and Betty Moore Foundation’s EPiQS Initiative through Grant No. GBMF4411. In addition, L.X. was funded, in part, by the W. M. Keck Foundation.

\bibliographystyle{apsrev}

\end{document}